\documentstyle[aps,prc]{revtex}
\date{\today}
\begin{document}
\flushbottom

\widetext
\draft
\title{First-Forbidden $\beta$ decay: \protect\\
Medium enhancement of the axial charge at A $\sim$ 16}
\author{E. K. Warburton}
\address{Brookhaven National Laboratory, Upton, New York 11973}
\author{I. S. Towner}
\address{AECL Research, Chalk River Laboratories, \protect\\
Chalk River, Ontario, Canada K0J 1J0}
\author{B. A. Brown}
\address{National Superconducting Cyclotron Laboratory and  \protect\\
Department of Physics and Astronomy, East Lansing, MI 48824}
\date{\today}
\maketitle

\def\thepage{\arabic{page}}
\makeatletter
\global\@specialpagefalse
\ifnum\c@page=1
\def\@oddhead{Draft\hfill To be submitted to Phys. Rev. C}
\else
\def\@oddhead{\hfill}
\fi
\let\@evenhead\@oddhead
\def\@oddfoot{\reset@font\rm\hfill \thepage \hfill}
\let\@evenfoot\@oddfoot
\makeatother

\begin{abstract}Calculations are presented for four relatively strong
first-forbidden $\beta$ decays in the region A=11-16 in order to
study the very large mesonic-exchange-current enhancement of the rank-zero
components. The $\mu^-$ capture on $^{16}$O is considered on the same footing.
The wave functions utilized include up to 4$\hbar\omega$ excitations.
Two-body exchange-current matrix elements are calculated as well as
one-body impulse approximation matrix elements. The resultant enhancement
factor that multiplies the
impulse-approximation axial-charge matrix element is thereby determined
by comparison to experiment to be
$\epsilon_{exp}$ = 1.61 $\pm$ 0.03 from three $\beta^-$
decays and $\mu^-$ capture on $^{16}$O, which is
in excellent agreement with
meson-exchange calculations in the
soft-pion approximation.
\\
{\bf PACS: {21.60.-n,21.60.Cs,23.40.Bw,21.10.Dr,27.20.+n,23.40.Hc}}
\end{abstract}

\makeatletter
\global\@specialpagefalse
\def\@oddhead{\hfill}
\let\@evenhead\@oddhead
\makeatother
\nopagebreak
\twocolumn
\narrowtext
\section{INTRODUCTION} \label{sec:intro}
This is one of several articles describing theoretical calculations of
first-forbidden $\beta$ decay observables and other related weak-interaction
variables with as high accuracy as is currently possible. Recent results have
been reported for A = 50 \cite{wa91a}, and the A=133-134 \cite{wa92c}, and
A=205-212 \cite{wa93c} regions. Here we report on the four A = 11-16 decays of
Fig. \ref{decay} and Table \ref{data}. The main motive
is to add understanding of the very large enhancement over the impulse
approximation of the rank-zero (R0) axial-charge matrix element $M^T_0$
observed
especially in the lead region \cite{wa91}.

The four $\beta$ decays of Fig. \ref{decay} are the known decays in light (A
$<$ 37) nuclei that are fast enough to provide potentially reliable information
on the medium enhancement of $M^T_0$. In addition to these four decays, an
important part of this study is a consideration of the inverse of
$^{16}$N(0$^-$) \raisebox{-.6ex}{$\stackrel{~\beta^-}{\longrightarrow}$}
$^{16}$O(0$^+$), namely $\mu^-$ capture on $^{16}$O
leading to the $0^-$ first-excited state of $^{16}$N with --- for the first
time --- a calculational precision comparable to that routinely used in
$\beta$-decay studies \cite{be82}.

The light nuclei have always been and will continue to be the premier testing
ground for our views on the structure of nuclei. Our main emphasis here will be
on A = 16 nuclei. $^{16}$O has a fascinating and complex structure
since
0$\hbar\omega$, 2$\hbar\omega$, and 4$\hbar\omega$ excitations are manifestly
apparent amongst the low-lying levels. The (0+2+4)$\hbar\omega$ model of
Brown and Green \cite{br66} was an early, successful, and important description
of these states. Recently, we described a large-basis shell-model
diagonalization of $^{16}$O in a (0+2+4)$\hbar\omega$ basis and $^{16}$N in a
(1+3)$\hbar\omega$ basis \cite{wa92b}. Similar wave functions will be
used in the $^{16}$N($0^-$)  $\leftrightarrow$  $^{16}$O($0^+$)
$\beta^-$ and $\mu^-$ calculations. The $\beta^-$ decays of $^{16}$C, $^{15}$C
and $^{11}$Be  will be treated in the more truncated (1+3)$\hbar\omega$
$\rightarrow$ (0+2)$\hbar\omega$ space (A = 11,15) or the
(2+4)$\hbar\omega$ $\rightarrow$ (1+3)$\hbar\omega$ space ($^{16}$C) with the
effects of 4$\hbar\omega$ or 5$\hbar\omega$ added perturbatively.

In Sec. II we give a review of the shell-model interactions used in
the present study.
In Secs. III.A and III.B we describe the calculation of the one-body (impulse
approximation) matrix elements, which enter in these rank-zero processes, and
how they are combined to give theoretical rates. The two-body (meson exchange)
matrix elements are considered in Sec. III.C.
In Sec. IV we describe the $^{16}$N($0^-$) $\leftrightarrow$ $^{16}$O($0^+$)
transitions and in Sec. V we take up the $\beta^-$ decays of $^{15}$C,
$^{11}$Be, and $^{16}$C, Finally, in Sec. VI we discuss the conclusions
concerning the enhancement factors for $\beta^-$ and $\mu^-$ capture.

\section{THE WAVE FUNCTIONS} \label{sec:wavefunctions}

Shell-model calculations were performed with the shell-model code
OXBASH \cite{br84}. With OXBASH,
spurious center-of-mass motion is removed by the usual method \cite{gl74} of
adding a center-of-mass Hamiltonian $H_{cm}$ to the interaction. The
shell-model
studies start with the recently constructed interactions of Warburton and
Brown \cite{wa92a}, which are based on interactions for
the 0p1s0d shells determined by a least-squares fit to 216 energy levels in the
A = 10-22 region assuming no mixing of n$\hbar\omega$ and (n+2)$\hbar\omega$
configurations. The 0p1s0d part of the interaction cited in Ref. \cite{wa92a}
as WBP results from a fit to two-body matrix elements (TBME) and
single-particle energies (SPE) for the p-shell and a potential representation
of the 0p-(1s0d) cross-shell interaction.
The WBP model space was expanded to include the 0s and 0f1p major shells
by adding the appropriate 0f1p and cross-shell 1s0d-0f1p two-body matrix
elements of the WBMB interaction \cite{wa90a} and all the other necessary
matrix elements from the bare G matrix potential of Hosaka, Kubo, and Toki
\cite{ho85}. The 0s, 0f, and 1p SPE were determined as described in
Ref. \cite{wa92a}. Thus the WBP interaction is constructed in a
similar manner to the Millener four-shell interaction described in
Ref. \cite{wa89}, but reproduces the binding energies of low-lying
$\ge$1$\hbar\omega$ levels in the A = 16 region with 2-3 times greater
accuracy.

Unless a complete model space is used for a given diagonalization, the
wave functions can contain spurious components. For a (0+2)$\hbar\omega$
calculation in $^{16}$O, the first four oscillator shells comprise a
complete basis, while for a (0+2+4)$\hbar\omega$ calculation the first
six shells must be included for completeness \cite{wa92b,ha90}. We began
our shell-model studies by diagonalizing the (0+2+4)$\hbar\omega$ $0^+$
T = 0 and (1+3)$\hbar\omega$ 0$^-$-3$^-$ T = 1 states of $^{16}$O in model
spaces comprising both the first six and first four oscillator shells.
As discussed in Ref. \cite{wa92b}, negligible difference was found in
the wave functions and observables of interest between the calculations
within these two model spaces. Thus the calculations reported here were
performed in the four-shell model spaces --- we emphasize that the
results in six-shell model spaces would be essentially identical.

In Ref. \cite{wa92b} several methods for constructing the mixed
$\hbar\omega$ spectra were explored, each of which gave a reasonable level
scheme for A=16. Since that work we have further explored the sensitivity of
other observables such as the electron scattering form factors from the
ground state of $^{16}$O to 0$^+$, 2$^+$, and 4$^+$ excited states and the
M1 decay of low-lying 1$^{+}$ T = 1 states \cite{br93}. We found that the
method
of lowering just the 4$\hbar\omega$ configurations for the positive-parity
states tends to produce too much mixing of the 4$\hbar\omega$
components into those states around 12-18 MeV, which are known ---
from experimental observables --- to be mostly 2$\hbar\omega$. The gap method
turns out to be better in this regard because it simultaneously lowers the
2$\hbar\omega$ configurations with respect to the 4$\hbar\omega$
configurations. The needed reduction of the 0p$-$1s0d gap --- initially
$\Delta$(0p-1s0d) = 11632 keV --- for the WBP
interaction is 3.0 MeV, which means that the 2$\hbar\omega$ states are shifted
down by 6.0 MeV and the 4$\hbar\omega$ states are shifted down by 12.0 MeV.
It is worth noting, as discussed in
Ref. \cite{wa92b}, that the amount that the 4$\hbar\omega$ configurations need
to be lowered is essentially equal to the shift of the ground state in
going from the 0$\hbar\omega$ space to the (0+2+4)$\hbar\omega$ space. Thus
the view can be taken that
the shifts are not free parameters but with a given $\hbar\omega$ truncation
can be determined self-consistently. When the negative-parity T = 1 states are
treated in a (1+3)$\hbar\omega$ space it is perhaps reasonable to shift the
3$\hbar\omega$ configurations down by 12.0 MeV and the
1$\hbar\omega$ configurations down by 6.0 MeV, and doing so
gives excellent agreement with experiment for the energies of
these states relative to the positive-parity T = 0 states. Note that this
is not the same as the pure gap method, which would involve shifts
of 9.0 and 3.0 MeV for the 3$\hbar\omega$  and
1$\hbar\omega$ configurations, respectively. Nevertheless we
will refer to this as the gap method since the structure of the
negative-parity states are exactly the same in either case (since
the structure only depends upon the 6.0-MeV shift difference between the
3$\hbar\omega$  and 1$\hbar\omega$ configurations).
We refer to the results obtained with the fitted potential for all matrix
elements and the gap method as WBP.

In addition, we have further explored the interaction
$V^{2\hbar\omega}$ that mixes n$\hbar\omega$ and (n+2)$\hbar\omega$
configurations. In Ref. \cite{wa92b} this interaction (40
0p-1s0d TBME) was generated from the potential obtained
from the fit described above (WBP). However, since the core-polarization
corrections may be different for 1$\hbar\omega$ and 2$\hbar\omega$
matrix elements, the use of a common potential
may not be entirely correct. As an alternative we have explored \cite{br93}
the use of the Bonn G matrix \cite{bo87} for the mixing interaction
$V^{2\hbar\omega}$. We find that the bare G matrix cannot reproduce the
energy-level spectrum $^{16}$O with any amount of 2$\hbar\omega$ and/or
4$\hbar\omega$ shift, but that the
interaction was acceptable in this regard
if it was renormalized by a factor
of 0.8. Such a renormalization seems reasonable because the
core-polarization corrections are found to reduce the average
$V^{2\hbar\omega}$ matrix elements of the bare G matrix
by about this amount \cite{kuo93}. We
will refer to results obtained with $V^{2\hbar\omega}$
replaced by 0.8 times the Bonn G matrix
as WBN. For this interaction the needed reduction in the 0p-1s0d gap was found
to be 3.5 MeV so the the 2$\hbar\omega$ and 4$\hbar\omega$ components are
lowered by 7 and 14 MeV, respectively. As for the WBP interaction, the spirit
of the gap method was retained for the odd-parity states in that the difference
between the 3$\hbar\omega$ and 1$\hbar\omega$  shifts was set at 7 MeV. However
the absolute shifts were arbitrarily set so as to minimize the difference in
the experimental and theoretical excitation energies of the low-lying T = 1
quartet. This, of course, has no influence on the wave functions of the states.

The main reason for exploring the use of the Bonn $V^{2\hbar\omega}$
interaction was
that the WBP interaction did not provide a satisfactory description of the
electron scattering form factors and of the M1 decay of the low-lying 1$^+$
T = 1 states.  The WBN interaction does considerably better in this regard
and, in fact, provides the most satisfactory description of these phenomena
of any realistic interaction that we are aware of \cite{br93}. Thus the results
given here for the WBN interaction are the preferred ones. The WBP results are
given to provide a measure of the sensitivity to the interaction used.

\section{THE WEAK INTERACTION PROCESSES}

\subsection{The one-body contribution} \label{sec:onebody}

Our concern here is with the one-body (impulse approximation) matrix elements
of the rank-zero (R0) axial current. These are the matrix elements of the R0
member of the spin-dipole
operator and of the helicity operator $\gamma_5$, which is commonly called the
time-like component of the axial current, or the axial charge. With $\gamma_5$
replaced by its non-relativistic limit (good to
order $1/M_N$) the single-particle
matrix elements (with the relative phase appropriate to
$\beta$ decay) are \cite{wa88a,footnote2,ed57,footnote4}
\begin{mathletters}
\label{allc}
\begin{eqnarray}
M^S_0(j_ij_f) &= &g_A\sqrt{3}\langle j_f
\Vert\vert ir[C_1,\mbox{\boldmath $\sigma$}]^0
\mbox{\boldmath $\tau$}\vert\vert\vert j_i\rangle C_{TJ} \label{c:1} \\
M^T_0(j_ij_f) &= -&g_A\sqrt{3}\langle j_f
\vert\vert\vert \case{i}/{M_N}[\mbox{\boldmath $\sigma,\nabla$}]^0
\mbox{\boldmath $\tau$}\vert\vert\vert j_i\rangle C_{TJ}
{\mathchar'26\mkern-9mu\lambda_{Ce}}^2, \label{c:2}
\end{eqnarray}
\end{mathletters}
where
\vspace{-0.2cm}
\begin{equation}
C_{TJ} =  \frac{(-1)^{T_f - T_{zf}}}{[2(2J_i+1)]^{1/2}}
\left( \begin{array}{c}
T_f ~~~~1 ~~~~~~T_i \\
-T_{zf} ~\Delta T_z ~~T_{zi}
\end{array}\right),
\label{cg:1}
\end{equation}
and
\vspace{-0.2cm}
\begin{equation}
C_1 = \sqrt{ \frac{4\pi}{3}} Y_1, \label{cg:2}
\end{equation}
and where ${\mathchar'26\mkern-9mu\lambda_{Ce}}$
--- the electron Compton wavelength divided by 2$\pi$ --- is incorporated into
Eq. (\ref{c:2}) so that both matrix elements have the dimensions of $fm$. The
$j_i$ and $j_f$ are a short-hand notation for all quantum numbers needed to
label the single-particle states.
In Eq. (\ref{cg:1}), the $J_i$ and $T$ dependencies result from the reduced
matrix elements in a spin and isospin space
and $\sqrt 2$ is from the definition of the isospin operator.
It is important to keep in mind
that the matrix elements of Eq. (\ref{allc}) are hermitian
conjugates, which for harmonic-oscillator radial wave functions results in the
identity
\begin{equation}
M^T_0(j_ij_f) =-\biggl\lbrack \frac{E_{osc}}{m_ec^2}\biggl\rbrack M^S_0(j_ij_f)
,~~~~~~~{\rm for~HO.}   \label{osc}
\end{equation}
The usual shell-model procedure is followed of combining these single-particle
matrix elements $M^\alpha_R(j_ij_f)$ with one-body transition
densities $D^{(1)}_R(j_ij_f)$ via
\begin{equation}
M^\alpha_R = \sum_{j_ij_f}{\cal M}^{\alpha}_R(j_ij_f) =
\sum_{j_ij_f}D^{(1)}_R(j_ij_f)M^\alpha_R(j_ij_f)  \label{mar}
\end{equation}
where the subscript $R$ denotes the rank of the operator
($R$ = 0 in this study).
The one-body transition densities given by
\begin{equation}
D^{(1)}_R(j_ij_f) =
\frac{\langle J_fT_f \vert \vert \vert [a^{\dag}_{j_i}
\otimes {\bar a}_{j_f}]^{\Delta T \Delta J}
\vert \vert \vert J_iT_i \rangle }
{[(2\Delta J + 1)(2\Delta T + 1)]^{\case{1}/{2}}} \label{OBTD}
\end{equation}
contain all the information on the initial and final many-body wave functions.
The $J_i T_i $ and $J_f T_f $ are a short-hand
notations for all quantum numbers
needed to describe the many-body wave functions and
$\Delta T$,$\Delta J$ are multipolarities of the one-body operator, which
in our case has $\Delta T = 1$ and $\Delta J = R = 0$.

In the previous calculations for heavier nuclei \cite{wa91a,wa92c,wa93c},
the single-particle matrix elements were augmented by multiplicative
renormalization factors $q_\alpha(j_ij_f)$ that represented first-order
core-polarization effects. In the present calculation the model space is
large enough to include all possible first-order core-polarizations, and so
the one-body operators appropriate to bare nucleons are used.

The $M^\alpha_R(j_ij_f)$ and also the the two-body matrix elements described
in Sec. III.C are calculated with a combination of
harmonic-oscillator (HO) and Woods-Saxon (WS) wave functions. The WS
parameters
were determined by extrapolation of results obtained by least-squares fits to
nuclear charge distributions for
$^{12}$C and $^{16}$O \cite{de87}. In this procedure the separation energies
were fixed as described below and the orbit occupancies were taken
from (0+2+4)$\hbar\omega$ wave functions for $^{16}$O \cite{wa92b} and $^{12}$C
\cite{br93}.
For HO wave functions, we first found values of $\hbar\omega$ for $^{12}$C and
$^{16}$O that reproduced the root-mean-square charge radii,
$\langle r^2 \rangle^{\frac{1}{2}}$, with these orbit occupancies.
For $^{16}$O, with
$\langle r^2 \rangle^{\frac{1}{2}}$ = 2.730 $fm$ \cite{de87}, the result is
$\hbar\omega$ = 13.60 MeV as opposed to 13.16 MeV, which is obtained assuming
an $^{16}$O closed shell. Values of $\hbar\omega$ for A = 11 and 15 were
obtained from the A = 12 and 16 values assuming a linear dependence on A.

The WS results depend on the separation energies $S(n)$ for the $\beta^-$
parent
state and $S(p)$ for the daughter. These are related by
\begin{equation}
S(p) - S(n) =  Q_\beta - 0.782 ~MeV. \label{snp2}
\end{equation}
where $Q_\beta$ $-$ 0.782 MeV is the difference in binding energies of the
initial and final states.
The separation energies for a particular common parent state with excitation
energy $E_x$ in the (A$-$1,Z$-$1) core are then
\jot0.5cm
\begin{mathletters}
\label{alls}
\begin{eqnarray}
&S(n) =& \Delta E_b(n) + E_x - E_i  \label{s:1} \\
&S(p) =& \Delta E_b(p) + E_x - E_f  \label{s:2}
\end{eqnarray}
\end{mathletters}
\noindent where $E_i$ and $E_f$ are the excitation energies of the initial and
final states in the (A,Z$-$1) and (A,Z) nuclei, respectively and
$\Delta E_b(n)$ is the difference in binding energies of the ground state of
the initial nucleus and the (A$-$1,Z$-$1) core nucleus, i.e. the neutron
separation energy for the ground states, and similarly for $\Delta E_b(p)$.

The main effect of small separation energies is to increase the relative
magnitude of the ``tail'' of the radial wave function. As the separation
energies increase, the difference between WS and HO wave functions lessens, and
becomes insignificant compared to our knowledge of the wave functions. Thus
we use HO wave functions for all single-particle transitions except those
allowed for $\nu (1s0d)$ $\rightarrow$ $\pi (0p)$; i.e., all others have large
effective values of $E_x$ and thus of S(n) and S(p).

The method developed by Millener \cite{mi85} and adopted to a similar
calculation \cite{wa91} for $^{206}$Tl $\rightarrow$ $^{206}$Pb was
used to estimate the effective value of $E_x$, $\langle E_x(j) \rangle$, to
use in Eq. (\ref{alls}). This method involves an inclusive calculation of the
spectroscopic amplitudes for neutron pickup from the initial state and proton
pickup from the final state to all possible core states and an evaluation of
the effective excitation energy of the core states from a consideration of
these amplitudes and the resulting dependence of the matrix elements on $E_x$.
In this determination all core-state excitation energies greater than 10 MeV
were fixed at 10 MeV. The results for
the rank-zero $\nu 1s_{1/2}$ $\rightarrow$ $\pi 0p_{1/2}$ and $\nu 0d_{3/2}$
$\rightarrow$ $\pi 0p_{3/2}$ transitions --- the two allowed $\nu(1s0d)$
$\rightarrow$ $\pi(0p)$ transitions --- are given in Table \ref{spn}.
The $\langle E_x(j) \rangle$ of Table \ref{spn} are calculated with the WBP
interaction with the lowest allowed n$\hbar\omega$ wave functions for each of
the three nuclei involved in each of the four cases. In contrast to these
simple wave functions, we present results in this study calculated for model
spaces as complex as (1+3)$\hbar\omega$ $\rightarrow$ (0+2+4)$\hbar\omega$. The
question might well be asked as to the relevance of Table \ref{spn} to these
more complex --- and hopefully more realistic --- calculations. To address
this question the procedure was applied to the $\nu 1s_{1/2}$ $\rightarrow$
$\pi 0p_{1/2}$ single-particle component in the $^{16}$N
$\rightarrow$ $^{16}$O rank-zero decay with typical (1+3)$\hbar\omega$
and (0+2+4)$\hbar\omega$ wave functions
for $^{16}$N($0^-$) and $^{16}$O($0^+$), respectively and
with typical (0+2)$\hbar\omega$ wave function for the $^{15}$N
$\case{1}/{2}^-$ states. In a (0+2)$\hbar\omega$ space, there are 265
$^{15}$N $\case{1}/{2}^-$ states, 41 of which are spurious.
The value of $\langle E_x(j) \rangle$ found for the 224 non-spurious
states is 32 keV, in close agreement with the value of zero keV associated
with the simple calculation of Table \ref{spn}.

For the $\nu 0d_{3/2}$ $\rightarrow$ $\pi 0p_{3/2}$ transitions
the large values of $\langle E_x(j) \rangle$ mean that the
matrix elements are not sensitive to its value, in fact, one might just as
well (in ignorance) use HO wave functions for the $j$ = $\case{3}/{2}$
transitions and we do so for the calculation of the one-body matrix elements
$M^S_0$ and $M^T_0$ and the two-body matrix element $M^\pi_\beta$ and
$M^\pi_\mu$.

\subsection{$\beta^-$ and $\mu^-$ rates} \label{sec:burates}

Note that all the numerical results given in this subsection are relevant to
$^{16}$N $\leftrightarrow$ $^{16}$O.
The relationship between experiment and theory is given here in a way that
displays the similarity between $0^-$ $\rightarrow$ $0^+$ $\beta$ decay
and  $0^+$ $\rightarrow$ $0^-$ $\mu^-$ capture, it follows closely the
treatments of Behrens and B\"uhring \cite{be82} for $\beta$ decay and Nozawa,
Kubodera, and Ohtsubo (NKO) \cite{no86} for both $\beta^-$ decay and
$\mu^-$ capture but especially the latter:
\begin{mathletters}
\label{alllam}
\begin{equation}
\Lambda_\beta
= \frac{G^2}{2\pi^3}\frac{f_0}
{{\lambdabar_{Ce}}^2}\vert M^{\beta}_0\vert^2, \\ \label{lam:1}
\end{equation}
\begin{equation}
\Lambda_\mu 
= C_{R} \frac{G^2}{2\pi}\frac{\omega^2}{1\! +\! \omega/M_f}\frac{\vert
{}~{\overline {\phi_{1s}({0})}}~ \vert^2}{{\lambdabar_{Ce}}^2}
\Biggl\lbrack
\frac{g_A(q^2)}{g_A(0)}\Biggl\rbrack^2 \!\vert M^\mu_0\vert^2. \label{lam:2}
\end{equation}
\end{mathletters}
\noindent All quantities are in natural units $\hbar$ = c = $m_e$ = 1. The
unit of time is the second and of length $\lambdabar_{Ce}$
 --- the electron Compton wavelength divided by 2$\pi$. The decay rate
$\Lambda$
($= 1/\tau$) has the units of sec$^{-1}$. G is the standard weak-interaction
constant. The $\beta$-decay phase-space factor is
\begin{equation}
f = \int^{W_0}_1 \!\!\!\!C(W)F(Z,W)(W^2-1)^{\case{1}{2}}W(W_0
\!-\! W)^2dW \label{f}
\end{equation}
where C(W) is the shape factor, F(Z,W) is the Fermi function, W is the
electron energy and W$_0$ the total disintegration energy --- both including
the rest mass. The allowed phase-space
factor, $f_0$ is given by the integral of Eq. (\ref{f}) with the shape factor
C(W) = 1. For a pure rank-zero (R0) decay as is involved here,
${\overline {C(W)}}$ = $f/f_0$ = $\vert M^{\beta}_0\vert^2$ where
the $M^{\alpha}_0$ ($\alpha \equiv \beta,\mu$) of Eq. (\ref{alllam}) are
combinations of matrix elements. We choose to give all matrix elements in $fm$;
thus $\lambdabar_{Ce}$ appears in Eq.
(\ref{alllam}). The axial-coupling constant $g_A(q^2)$ is a
function of the four-momentum transfer $q$. For nuclear $\beta$ decay
$q = 0$ to a good approximation. We incorporate $g_A(0)$
($\equiv$ $g_A$) into the matrix elements. Hence it does not appear in Eq. (9a)
and normalizes the $q$-dependent ratio $g_A(q^2)$
in Eq. (\ref{lam:2}). In Eq. (\ref{lam:2}), $C_{R}$ is a second-order
relativistic correction (see Appendix A), $\omega$ is the muon-neutrino
energy, $(1+\omega/M_f)^{-1}$ is a recoil correction with $M_f$ being the
mass of the final nucleus, and
${\vert \overline{\phi_{1s}({\bf r})} \vert}^2_{r = 0}$ is the
probability of finding the $\mu^-$ at the origin (see Appendix A):
\begin{equation}
{\vert \overline{\phi_{1s}({0})} \vert}^2 = {\cal R}^2_Z{\vert
\phi_{1s}(0) \vert}^2_{{\rm\scriptstyle point}\atop{\rm\scriptstyle nucleus}}
 = {\cal R}^2_Z \frac{(Z\alpha m^r_\mu)^3}{\pi}. \label{phi0}
\end{equation}
In Eq. (\ref{phi0}),
$m^r_\mu$ is the reduced $\mu^-$ mass and ${\cal R}_Z$ is a correction factor
obtained by solving the Dirac equation for the wave function of the muon in
the field of a finite-charge distribution (see Appendix A).
Parameter values (in natural units unless otherwise specified)
relevant to $^{16}$N($0^-$) $\leftrightarrow$ $^{16}$O($0^+$) are given
in Table \ref{def}. Using these parameters we find
\begin{mathletters}
\label{allla}
\begin{eqnarray}
\Lambda_\beta &= 1.4206\cdot10^{-4}&\vert M^\beta_0 \vert^2
= 0.485 \pm 0.019 ~sec^{-1}, \label{la:1} \\
\Lambda_\mu   &= ~~~~~~~0.1210&\vert M^\mu_0 \vert^2
= 1560 \pm 94 ~sec^{-1} \label{la:2}
\end{eqnarray}
\end{mathletters}
where the experimental results on the right are taken from
Refs. \cite{ha85} and \cite{to86}.
The experimental rates lead to the following experimental matrix elements:
\begin{mathletters}
\label{allmexp}
\begin{eqnarray}
M^\beta_0 &= 83.90[\Lambda_\beta(s^{-1})]^{\case{1}/{2}}=~~58.4& \pm 1.1 ~fm,
\\
\label{mexp:1}
M^\mu_0 &= 2.874[\Lambda_\mu(s^{-1})]^{\case{1}/{2}}   = 113.5& \pm 3.4 ~fm.
\label{mexp:2}
\end{eqnarray}
\end{mathletters}
Results for the $\beta$-decay matrix elements for the other
three cases of interest are given in Table \ref{data}.
It is these experimental matrix elements that we will compare to theory.
We write $M^\beta_0$ and $M^\mu_0$ in the form

\begin{eqnarray}
M^\beta_0 &= [M^T_0 + M^\beta_\pi + a^{\beta}_SM^S_0]
&= [M^T_0 \epsilon^{\beta} + a^{\beta}_SM^S_0]~~~ \label{allmth} \\
M^\mu_0 &= [a^\mu_TM^T_0 + M^\mu_\pi - a^{\mu}_SM^S_0]
&= [a^\mu_TM^T_0 \epsilon^{\mu} - a^{\mu}_SM^S_0], \nonumber
\end{eqnarray}
\noindent where the $M^\alpha_\pi$ are the two-body meson-exchange current
(mec) matrix elements defined in Sec. III.C below. When Eqs. (\ref{allmexp})
and (\ref{allmth}) are used to compare experiment and theory, it is
conventional
to extract values of $\epsilon^\alpha$ that are required to reproduce
experiment, and these will be referred to as $\epsilon^\alpha_{exp}$.
The calculated values for $\epsilon^\alpha$ based upon these equations
together with the calculated two-body
$M^\alpha_\pi$ discussed in Sec \ref{sec:twobody} will be referred to as
$\epsilon^\alpha_{mec}$.

Our evaluation of the $\beta$-decay rate follows the rigorous and accurate
treatment of Behrens and B\"uhring \cite{be82}. In writing the expression for
$M^\beta_0$ in Eq. (\ref{allmth})
small terms included in the first-order treatment have been neglected. However,
for the decays in question these terms contribute less than 0.2\% to
$M^\beta_0$. 
In the Behrens-B\"uhring treatment the $a^\beta_S$ of
Eq. (\ref{allmth}) is given by \cite{be82,wa88a,to81,footnote5}
\begin{equation}
a^\beta_S = \case{1}/{3}(Q_\beta + 1) + \xi r^\beta_S = 7.208 + 3.187r^\beta_S,
\label{as}
\end{equation}
where $Q_\beta$ is dimensionless (in units of the electron mass) and the
numerical results apply to the $^{16}$N $\rightarrow$ $^{16}$O decay. Note the
$r^\beta_S$ and the $a^\alpha_S$ and $a^\alpha_T$ of Eqs. (14) and (15) are
positive-definite quantities so that
the contributions of $M^T_0$ and $M^S_0$ add destructively in forming
$M^\beta_0$ and constructively in forming $M^\mu_0$.

The $a^\mu_S$ and $a^\mu_T$ of Eq. (\ref{allmth}) and $r^\beta_S$ of
Eq. (\ref{as})
are defined as ratios of matrix elements evaluated with extra radial factors
to the normal matrix elements. The radial dependencies for $\mu^-$ capture are
given explicitly in Appendix A.
In Eq. (\ref{as}) the ratio $r^\beta_S$ is
insensitive to the wave functions used. Approximate values of $a^\beta_S$
are $\approx$ 7.80, 8.52, 7.45 and 9.40 for the decays of $^{11}$Be, $^{15}$C,
$^{16}$C and $^{16}$N respectively. Note however that all
$a^\alpha_T$ and $a^\alpha_S$ are calculated explicitly for the wave functions
at hand.

\subsection{The two-body soft-pion contribution} \label{sec:twobody}

Towner \cite{to92} has  recently made an investigation of the one-pion
exchange contribution to single-particle matrix elements of the axial charge
in (closed shell $\pm$ 1)
nuclei from A $\sim$ 16 to 208. Towner's results indicate that for A $\sim$ 16
the soft-pion diagram \cite{ku78,gu78} alone gives an adequate
representation of the meson exchange and so we only consider this term.
The soft-pion contribution was incorporated into the shell-model calculations
in the manner described for the similar parity-nonconserving (PNC) matrix
element \cite{br80}. The general expression for the soft-pion
contribution is a sum of the product of a two-body transition density
$D^{(2)}_0(j_1j_2JTj_3j_4JT^\prime)$
and a two-body meson-exchange matrix element:

\begin{eqnarray}
&&M^\beta_\pi = G_{soft \pi}C_{TJ}
\sum_{{j_1 \le j_2}\atop{{j_3 \le j_4}\atop{JTT^\prime}}}
- D^{(2)}_0(j_1j_2JTj_3j_4JT^\prime)\times  \label{pi} \\
\vspace{-0.2cm}
&&\langle j_1 j_2 JT\vert \vert \vert{\hat g}(r_r)
i\mbox{\boldmath $(\sigma_1 + \sigma_2)$}\cdot
{\hat {\mbox{\boldmath $r$}}}_r
\mbox{\boldmath$ (\tau_1 \times \tau_2)$}Y_\pi(x_\pi)
\vert \vert \vert j_3 j_4 JT^\prime \rangle \nonumber
\end{eqnarray}
where $x_\pi$ =
$m_\pi r_r$ with \mbox{\boldmath $r_r$} = \mbox{\boldmath $r_1 - r_2$},
$Y_\pi(x_\pi) = (1 + 1/x_\pi)e^{-x_\pi}/x_\pi$ and ${\hat g}(r_r)$ is a
short-range correlation (SRC) function. The SRC used in this work was
${\hat g}(r_r)$ = $1 - j_0(q_cr_r)$ with $q_c$ = 3.93 $fm^{-1}$ \cite{to92}.
We will also compare results obtain with this SRC and with the simple
cut-off factor $\theta(r_r - d)$ for which ${\hat g}(r_r)$ = 1 for $r_r$ $>$ d
and 0 for $r_r$ $\le$ d. The soft-pion coupling constant is defined as
\begin{equation}
G_{soft \pi} = \frac{\sqrt 2}{8\pi }\frac{g^2_{\pi NN}}{g_A}
\frac{m^2_{\pi}}{M^2_N}
\mathchar'26\mkern-9mu\lambda_{Ce}
 = 69.74 ~fm \label{gsp}
\end{equation}
where we have used $g_{\pi NN} = 13.684$
\cite{bo87}.
The two-body transition density of Eq. (\ref{pi}) is given by
\begin{eqnarray}
&&D^{(2)}_0(j_1j_2JTj_3j_4JT^\prime) = \label{tbdt} \\
\vspace{-0.2cm}
&&\frac{\langle J_fT_f \vert\vert\vert
\lbrace \lbrack {a^{\dag}_{j_1}}\!\otimes\! {a^{\dag}_{j_2}}\rbrack^{JT}
\!\otimes\! \lbrack {{\bar a}_{j_3}}\!\otimes\!
{{\bar a}_{j_4}}\rbrack^{JT^\prime}
\rbrace^{\Delta J = 0,\Delta T = 1}
\vert\vert\vert J_iT_i \rangle}{\lbrack (2\Delta J + 1)(2\Delta T + 1)
(1 + \delta_{j_1j_2})(1 + \delta_{j_3j_4}) \rbrack^{1/2}}. \nonumber
\end{eqnarray}
As for the other shell-model calculations, the evaluation of $M^\beta_\pi$
involved four oscillator shells and the use of mixed HO and WS radial
wave functions. Our results for single-particle
transitions are identical to the soft-pion results of Towner \cite{to92}. Here
we consider more complicated transitions. The computer
program used is formulated in terms of harmonic-oscillator
wave functions. Our method of allowing Woods-Saxon radial wave functions
is to expand the appropriate WS radial wave function in terms of HO wave
functions (up to 10 terms). As explained in Sec. III.A, we use WS wave
functions for the $\nu 1s_{1/2}$ $\rightarrow$ $\pi 0p_{1/2}$ transition but
use HO wave functions to evaluate the rest. As for the one-body matrix element,
$\mu^-$ capture differs from $\beta$ decay in that the analogous $M^\mu_\pi$ to
the $M^\beta_\pi$ of Eq. (\ref{pi})
should be evaluated with the factor $a^\mu_T$ [$\approx j_0(\omega r)$] of Eq.
(A17) of the Appendix inserted. Thus a dependence on
{\boldmath$r$} [$\equiv$ $\case{1}/{2}${\boldmath$(r_1 + r_2)$}] as well as on
{\boldmath $r_r$} is introduced. This dependence is approximated by retaining
terms up to order $\omega^2$  with the result that the replacement
\begin{eqnarray}
&&\mbox{\boldmath $(\sigma_1 + \sigma_2)$}\cdot
{\hat {\mbox{\boldmath $r$}}}_r \rightarrow
\mbox{\boldmath $(\sigma_1 + \sigma_2)$}\cdot
{\hat {\mbox{\boldmath $r$}}}_r \nonumber \\
\vspace{-0.1cm}
&&- \frac{1}{6}\omega^2(r^2 + \frac{1}{4}r^2_r)
\mbox{\boldmath $(\sigma_1 + \sigma_2)$}\cdot
{\hat {\mbox{\boldmath $r$}}}_r \label{2j0} \\
\vspace{-0.1cm}
&&+ \frac{1}{6}\omega^2(\mbox{\boldmath $r\cdot r$}_r)
\mbox{\boldmath $(\sigma_1 - \sigma_2)$}\cdot
{\hat {\mbox{\boldmath $r$}}}_r \nonumber
\end{eqnarray}
was made in Eq. (\ref{pi}) to obtain $M^\mu_\pi$ \cite{footnote6}.

The relative contributions of $M^\beta_\pi$ to $M^T_0$ is formulated in terms
of the $\epsilon^\beta$ parameter defined in Eq. (\ref{allmth}). It is
informative to
consider the values of $\epsilon^\beta_{mec}$ obtained for simple
configurations. Evaluations with HO and WS wave functions are collected in
Table \ref{ist} for
two different SRC factors. It is seen that $\epsilon^\beta_{mec}$ has
a fairly strong state dependence. Because of this state
dependence and because the two-body and one-body matrix elements have
different dependencies on the nuclear wave functions, Eq. (\ref{pi}) or an
equivalent expression must be used for a rigorous evaluation of
$\epsilon^\beta_{mec}$. However, a useful approximation is
\begin{equation}
\epsilon^\beta_{mec} \approx
\frac{\sum_{j_ij_f}D^{(1)}_R(j_ij_f)\epsilon^\beta_{mec}(j_ij_f)M^T_0(j_ij_f)}
{\sum_{j_ij_f}D_R(j_ij_f)M^T_0(j_ij_f)}, \label{approx}
\end{equation}
with a similar expression for $\epsilon^\mu_{mec}$.
Comparison shows that this approximation --- used with the results listed
in Table \ref{ist} to represent the $\epsilon^\beta_{mec}(j_ij_f)$ ---
underestimates $\epsilon^\beta_{mec}$ by $\sim$5\%. Nevertheless, this
approximation is useful because the evaluation of
Eq. (\ref{pi}) takes an unusually large amount of computer time. In comparisons
of different diagonalizations and tests of different radial forms or SRC
factors
we make use of Eq. (\ref{approx}).

\section{RESULTS FOR $^{16}$N($0^-$) $\leftrightarrow$ $^{16}$O($0^+$)}
\label{sec:16bu}

As a first orientation, consider the four decays of Table \ref{data} in
successive degrees of
model-space complexity. The simplest of these is the ``single-particle''
approximation in which only the $0p_{1/2}$ and $1s_{1/2}$
degrees of freedom are allowed. Next allow for mixing within a single
major-oscillator shell but keep only the lowest possible $\hbar\omega$
configurations, e.g., 1$\hbar\omega$ $\leftrightarrow$ 0$\hbar\omega$
for the $^{16}$N $\leftrightarrow$ $^{16}$O transitions. In
this approximation the WBP and WBN interactions are equivalent. The results
for the two calculations are compared in Table \ref{simple}.
First we observe that in all cases there is a significant
reduction in going from the simple configuration to the major-oscillator
configuration. This is primarily due a destructive interference between
the $\nu 1s_{1/2}$ $\leftrightarrow$ $\pi 0p_{1/2}$ and $\nu 0d_{3/2}$
$\leftrightarrow$ $\pi 0p_{3/2}$ terms in Eq. (\ref{mar}).

The next level of approximation allows for those terms which
enter in the next order in a perturbation expansion.
For 1$\hbar\omega$ $\rightarrow$ 0$\hbar\omega$ transitions, this means
including 2$\hbar\omega$ admixtures in the final state since these connect
directly to the 1$\hbar\omega$ initial state; 3$\hbar\omega$ admixtures in the
initial state are not included in this order
because they do not directly connect to the 0$\hbar\omega$ final state and
it is inconsistent to include them without also including 4$\hbar\omega$ terms
in the final state. A consistent higher-order calculations involves
(1+3)$\hbar\omega$ $\rightarrow$ (0+2+4)$\hbar\omega$ transitions and
is only possible at present for $^{16}$N decay, which we now consider.

The model-space dimensions for $^{16}$N 0$^-$ in a (1+3)$\hbar\omega$ model
space and $^{16}$O 0$^+$ in a (0+2+4)$\hbar\omega$ model space are 713 and
4255,
respectively.
The gap reduction for the WBP interaction is 3.0 MeV. The lowest five
$^{16}$O 0$^+$ T = 0 states are calculated to lie at
0.00, 6.59, 10.77, 11.69 and 14.25 MeV with the 6.59-MeV
state predominately (88\%) 4$\hbar\omega$. The structure of the ground state is
43.2 percent 0$\hbar\omega$, 43.2 percent 2$\hbar\omega$ and
13.6 percent 4$\hbar\omega$. The lowest two 0$^-$ T = 1 states occur at
12.38 and 17.37 MeV and the structure of the lowest 0$^-$ and 2$^-$ T = 1
states are both (coincidentally) 71.1 percent 1$\hbar\omega$ and 28.9 percent
3$\hbar\omega$.

A partial WBN level scheme is compared to experiment in Fig. \ref{16}.
The gap reduction for the WBN interaction is 3.5 MeV.
The structure of the ground state is 34.9 percent 0$\hbar\omega$,
46.7 percent 2$\hbar\omega$ and 18.4 percent 4$\hbar\omega$.
The 6.44-MeV 0$^+_2$ state is predominantly (87\%) 4$\hbar\omega$.
The structure of the lowest 0$^-$ T = 1 state is
64.5 percent 1$\hbar\omega$ and 35.5 percent 3$\hbar\omega$.

The $\beta^-$ R0 and R2 decays of the $0^-$ and $2^-$ states of $^{16}$N are
illustrated in Fig. \ref{16N}.
Results for the $^{16}$N($0^-$) $\rightarrow$ $^{16}$O($0^+$) R0 $\beta$
decay are given in Table \ref{highbe} and those for the
$^{16}$O($0^+$) $\rightarrow$ $^{16}$N($0^-$) R0 $\mu^-$ capture are
given in Table \ref{highmu}. As a further test of the wave functions,
we compare with experiment \cite{wa85,wa68}
in Table \ref{highR2} R2 results for the
$^{16}$N($2^-$) $\rightarrow$ $^{16}$O($0^+$) unique first-forbidden
$\beta$ decays, which are simpler and better understood \cite{wa92b,wa88a}
than the R0 decays. For example, no appreciable mesonic contribution to R2
transitions is expected. As in Ref. \cite{wa92b} the good
agreement between theory and experiment for these R2 transitions
provides some evidence for the correctness of our wave functions.

The decompositions of M$^T_0$ and M$^{\textstyle z}_2$ obtained with the
WBP interaction for the R0 and R2 $\beta$ decays to the $^{16}$O ground state
into the four possible n$\hbar\omega$ $\rightarrow$
(n$\pm$1)$\hbar\omega$ components are shown in Fig. \ref{schematic}. The
various n$\hbar\omega$ contributions follow the classic
pattern found for de-excitation of E1-like particle-hole
configurations in a previous study of first-forbidden decays
in the A = 40 region \cite{wa88a}. In particular, the
1$\hbar\omega$ $\rightarrow$ 0$\hbar\omega$
and 3$\hbar\omega$ $\rightarrow$ 2$\hbar\omega$
matrix elements and the 1$\hbar\omega$ $\rightarrow$ 2$\hbar\omega$
and 3$\hbar\omega$ $\rightarrow$ 4$\hbar\omega$ matrix elements are closely
equal and the latter two are out of phase with the former. The detailed
composition of the ${\cal M}^\alpha_0$ of Eq. (\ref{mar}) for the
$^{16}$N(0$^-$) $\rightarrow$ $^{16}$(0$^+$) $\beta^-$ decay is given in Table
\ref{calm}.

\section{RESULTS FOR $^{15}$C, $^{11}$B\lowercase{e}, and $^{16}$C $\beta^-$
Decay} \label{sec:allbeta}

In this section we will only consider results obtained with the WBN interaction
using the gap method and mixed WS and HO single-particle radial wave functions.
We will confine our comparison of experiment and theory to a quotation of the
M$^\alpha_0$ and the value of $\epsilon^\beta_{exp}$, which when combined with
the calculated M$^\alpha_0$ reproduces experiment.  In all three cases, the
calculated $\epsilon^\beta_{mec}$ differ negligible from that calculated for
$^{16}$N decay.

\subsection{$^{15}$C($\case{1}/{2}^+$) $\rightarrow$
$^{15}$N($\case{1}/{2}^-$)}

The dimension D(J=$\case{1}/{2}$) of the J-matrix for the T = $\case{3}/{2}$,
$J^\pi$ = $\case{1}/{2}^+$ states of $^{15}$C is 2369 in the four-shell
(1+3)$\hbar\omega$ model space. For the T = $\case{1}/{2}$, $J^\pi$ =
$\case{1}/{2}^-$ states of $^{15}$N D($\case{1}/{2}$) is 23762 for a
(0+2+4)$\hbar\omega$ calculation and 265 for a (0+2)$\hbar\omega$
calculation. We can easily manage the $^{15}$C (1+3)$\hbar\omega$
and $^{15}$N (0+2)$\hbar\omega$ calculations but the (0+2+4)$\hbar\omega$
calculation is beyond our present resources. An attempt to truncate the
$^{15}$N (0+2+4)$\hbar\omega$ calculation by restricting the 4$\hbar\omega$
part to 4p-4h (4 particle--4 hole) excitations between the 0p and 1s0d
shells failed because the low-lying states in this truncation were highly
spurious. This is in contrast to similar calculations performed for the
$0^+$ states of $^{16}$O \cite{wa92b} and $^{12}$C \cite{br93}. (The
low-lying $0^+$ states in the latter calculations were nearly spurious free.)
However, the importance of
the 4$\hbar\omega$ component in the $^{16}$O ground state suggests a similar
importance in $^{15}$N and we would like to include an estimation of its
effect.
Thus a (1+3)$\hbar\omega$ $\rightarrow$ (0+2)$\hbar\omega$ calculation was made
with the effects of adding a 4$\hbar\omega$ term to $^{15}$N estimated by
assuming the same amplitude of 4$\hbar\omega$ in the $^{15}$N ground state as
in $^{16}$O (see Fig. 4) and assuming
\begin{equation}
M(3\hbar\omega \rightarrow 4\hbar\omega) \approx
\frac{M(1\hbar\omega \rightarrow 2\hbar\omega)}
{M(1\hbar\omega \rightarrow 0\hbar\omega)}
M(3\hbar\omega \rightarrow 2\hbar\omega) \label{m}
\end{equation}
where M[(n $\pm$ 1)$\hbar\omega$ $\rightarrow$ n$\hbar\omega$] is the
specific component, such as in Fig. 4, of either $M^T_0$ or $M^S_0$.
Thus Eq. (\ref{m}) quantifies our observations on the systematics of Fig. 4.

A value of $\Delta_{3\hbar\omega}$ = $-$8.50 MeV was determined for the
$^{15}$C (1+3)$\hbar\omega$ calculation by a least-squares matching to the
eleven experimental even-parity A = 15 T = $\case{1}/{2}$ energy levels for J
$\le$ $\case{5}/{2}$ states below 11 MeV, the yrast T = $\case{3}/{2}$
$\case{1}/{2}^+$, and $\case{5}/{2}^+$ states, and the yrast T = $\case{1}/{2}$
$\case{13}/{2}^+$ state. The value of $\Delta_{2\hbar\omega}$ used in the
$^{15}$N calculation was determined as $-$8.30 MeV from a similar consideration
of the A = 15 odd-parity spectrum.

The results are M$^T_0$ = 27.66 $fm$, M$^S_0$ = 1.53 $fm$,
$\epsilon^\beta_{exp}$ = 1.54 $\pm$ 0.04 (uncertainty from experiment only).

\subsection{$^{11}$Be($\case{1}/{2}^+$) $\rightarrow$
$^{11}$B($\case{1}/{2}^-$)}

The calculation for ${11}$Be($\case{1}/{2}^+$) $\rightarrow$
${11}$B($\case{1}/{2}^-$) follows the procedure just described for
${15}$C($\case{1}/{2}^+$) $\rightarrow$ ${15}$N($\case{1}/{2}^-$) with
the reliance on the (0+2+4)$\hbar\omega$ $^{16}$O $0^+_1$ wave function in the
latter case changed to a similar reliance on the (0+2+4)$\hbar\omega$ $^{12}$C
$0^+_1$ wave function. The D($\case{1}/{2}$) for the (1+3)$\hbar\omega$
$\rightarrow$ (0+2)$\hbar\omega$ calculation are 5674 and 1063, respectively.
Little of a definite nature is known about the 2$\hbar\omega$ and
3$\hbar\omega$
states of $^{11}$B and $^{11}$Be. Thus the values of $\Delta_{2\hbar\omega}$
and $\Delta_{3\hbar\omega}$ used in the calculations were assumed to be equal
and were obtained by linear interpolation between values found for $^{10}$B,
$^{12}$C and $^{15}$N. The resulting value is $-$4.00 MeV.

The effect of the 4$\hbar\omega$ component in $^{11}$B was considerably less
than the similar effect in $^{15}$N simply because the $^{12}$C ground state
wave function has only 1.86\% 4$\hbar\omega$ as opposed to 13.6\% in $^{16}$O.

The results are M$^T_0$ = 14.05 $fm$, M$^S_0$ = 1.10 $fm$,
$\epsilon^\beta_{exp}$ = 1.56 $\pm$ 0.08 (uncertainty from experiment only).

\subsection{$^{16}$C($0^+$) $\rightarrow$ $^{16}$N($0^-$)}

After considering $^{16}$C(0$^+$) $\rightarrow$ $^{16}$N(0$^-$) in detail we
conclude that the calculated matrix elements are unusually sensitive to the
details of the calculation and thus not suitable for a determination of
$\epsilon^\beta_{exp}$. To understand this, first consider the simple
2$\hbar\omega$ $\rightarrow$ 1$\hbar\omega$ calculation of Table \ref{simple}.
In the case of $^{16}$C the reduction of the M$^\alpha_0$ in going from
the simple single-particle configuration to the major-oscillator configuration
is unusually large and interaction dependent, because of the competition
between
the $0d_{5/2}^2$ and $1s_{1/2}^2$ configurations in the $^{16}$C ground state.
As a consequence of this,
the configuration in $^{16}$C, which is responsible for the destructive
$\nu 0d_{3/2}$ $\rightarrow$ $\pi 0p_{3/2}$ term (but which accounts for only
about 0.3 percent of the $^{16}$C wave function), results in a $\sim$50\%
reduction of the matrix elements.

Now consider the mixing of 2$\hbar\omega$ and 4$\hbar\omega$ components.
A (2+4)$\hbar\omega$ calculation --- which has D(0$^+$) =
4055 --- was performed.
The same decrease of the 0p-1s0d energy gap, 3.5 MeV,
for the T = 2 $0^+$ states was used as was applied for the T = 0
$0^+$ states. We find the first 10 states (i.e., all that were
examined) to be $>$44\% 4$\hbar\omega$ with, e.g., the ground state
being 55\% 2$\hbar\omega$ and 45\% 4$\hbar\omega$. The ($0^+_1,2$) excitation
energy in $^{16}$O was 22.2 MeV as compared to the experimental value of 24.52
MeV. The prominance of the 4$\hbar\omega$ component may seem surprising at
first sight; however, if a weak-coupling approximation such as that of Bansal
and French \cite{ba64} is used to
estimate the binding energies of $^{14}$C$\otimes$$^{18}$O, and
$^{12}$C$\otimes$$^{20}$O, one finds that these energies are nearly
degenerate so that which lies lowest and the energy gap between them depends
on the details of the interaction. Thus it is not surprising that the
(2+4)$\hbar\omega$ mixing is large and unusually sensitive to the details of
the calculation.

Because of these sensitivities we believe that the $^{16}$C decay is
primarily a test of the wave functions and does not provide a good measure of
the mesonic-exchange-current enhancement. This extreme wave function
sensitivity for $^{16}$C also
applies to similar R0 decays of $^{17}$N  and $^{18}$Ne \cite{mi85}.
We do not consider this decay any further in this study.

\section{DISCUSSION AND SUMMARY} \label{sec:conclusions}

We comment first on the value of the enhancement factor for time-like
axial-charge matrix elements deduced from a comparison of the impulse
approximation with experiment, $\epsilon^\alpha_{exp}$.
It is evident from Tables \ref{highbe}, and \ref{highmu}
that the value of $\epsilon^\alpha_{exp}$ is strongly dependent on the
single-particle radial wave function used to describe the $\nu 0s_{1/2}$
$\rightarrow$ $\pi 0p_{1/2}$ transition. The WS result is strongly preferred
since it provides the most realistic estimate of these radial wave functions.
The HO results are included to give some indication of the sensitivity of the
results to the radial form. We will not consider the HO results further.
Likewise, the results of the WBN interaction are strongly preferred over those
of the WBP interaction for the reasons stated in Sec. II. Again, the WBP
results are listed to give an indication of the sensitivity to the
shell-model interaction and will not be considered further.

In addition to the dependence on single-particle wave functions and
shell-model interactions, there is a further dependence we have not yet
considered; namely, the dependence of the $\mu^-$ capture rate on the
pseudoscalar coupling constant $g_p$ (see Appendices A and B).
For the WBN interaction with WS wave functions the dependence on $g_p$ can be
expressed as
\begin{equation}
\epsilon^\mu_{exp} = 1.55 + 0.085(g_p - 6.939). \label{gp}
\end{equation}

\noindent In the evaluation of the $\beta^-$ R0 matrix element $M^\beta_0$ the
contributions of $M^T_0$ and $M^S_0$ add destructively while they add
constructively in $M^\mu_0$ [see Eq. (\ref{allmth})]. This relative phasing
results in
less model-dependence for the determination of $\epsilon^\mu_{exp}$ than for
that of $\epsilon^\beta_{exp}$ and it has often been asserted that, for this
reason, $\mu^-$ capture provides a considerably more reliable determination of
$\epsilon_{exp}$ than $\beta$ decay. However, the uncertainty in $g_p$
(see Appendix B) completely negates this conclusion; from Eq. (\ref{gp}) we
see that the range of $g_p$ of $\sim$7-12 allowed by analysis of radiative
$\mu^-$ capture (see Appendix B) corresponds to a range in $\epsilon_{exp}$
values of
1.55-2.00 --- a range large enough to offset the afore-mentioned advantage.

Our adopted value for $\epsilon_{exp}$ follows from a consideration of the
results for the three $\beta^-$ decays that were analyzed. The average of
the three results (1.54 $\pm$ 0.04, 1.56 $\pm$ 0.08, 1.63 $\pm$ 0.02) quoted
for $\epsilon_{exp}$  is 1.60 $\pm$0.03 and this is
the result we shall adopt. We note that the $\mu^-$ capture result of 1.55
$\pm$ 0.08 (for the PCAC value of $g_p$) is in good accord with this value
especially if the small difference between $\epsilon^\beta_{exp}$ and
$\epsilon^\mu_{exp}$ of $\sim$0.02 is added to the $\epsilon^\mu_{exp}$ value.

The previous most ambitious calculation of the processes in question were the
(1+3)$\hbar\omega$ $\rightarrow$ (0+2)$\hbar\omega$ calculation of
Warburton \cite{wa88b} for the four $\beta^-$ decays of Table I and the
(1+3)$\hbar\omega$ $\leftrightarrow$ (0+2+4)$\hbar\omega$ results of
Haxton and Johnson \cite{ha90} for the $^{16}$N(0$^-$) $\leftrightarrow$
$^{16}$O(0$^+$) $\beta^-$ and $\mu^-$ processes. The calculations of Warburton
\cite{wa88b}, using the MK interaction, result in an average
$\epsilon^\beta_{exp}$ of 1.64 for the decays of Table I, in good agreement
with the present result of 1.61. Haxton and Johnson obtained results very close
to the present results given for the WBP interaction with WS wave
functions \cite{footnote7}. This is not surprising to us since we had  found
that
the $V^{2\hbar\omega}$ derived from the Kuo bare G matrix \cite{kuo93} ---
the $V^{2\hbar\omega}$ of the MK interaction used by Haxton and Johnson ---
has quite similar properties to the $V^{2\hbar\omega}$ of the WBP potential.
It is the change to the Bonn potential for $V^{2\hbar\omega}$ that gives
us a substantial improvement in the agreement of experiment and theory.

The theoretical values for the enhancement factors, $\epsilon^\beta_{mec}$ and
$\epsilon^\mu_{mec}$,
calculated from a meson-exchange model in the soft-pion approximation are very
close to our adopted experimental value and show
very little model dependence varying from 1.60 to 1.63 and 1.59 to 1.61,
respectively, depending on the choice of interaction and single-particle
wave functions.  This insensitivity has been noted previously
in work on parity non-conserving interactions \cite{ah85}
and follows because the
two-body matrix element can be represented quite well by an effective one-body
matrix element proportional to $\boldmath \sigma \cdot p$
--- and thus to $M^T_0$ ---
consequently the ratio of
the two-body to one-body matrix elements is not sensitive to nuclear structure.
However, as discussed by
Towner \cite{to92}, the results from meson-exchange models are dependent on
the choice of a short-range correlation function.
This is particularly true for the short-range operators discussed in
\cite{to92} originating in heavy-meson exchange.  In this paper, we
have only considered the long-range pion-exchange operators evaluated
in the soft-pion approximation for which the sensitivity to SRC is
somewhat less.  Some sample
calculations are given in Table IV. For the all-important $\nu 1s_{1/2}$
$\rightarrow$ $\pi 0p_{1/2}$ transition the use of the $\theta(r_r - 0.71)$
form reduces $\epsilon^\alpha_{mec}$ by a factor 0.96 compared to the
$1 - j_0(3.93r_r)$ form and the other transitions in this table have similar
ratios. In the full calculation for both $\beta$-decay and $\mu^-$ capture
($\alpha \equiv \beta,\mu)$, the reduction factor is 0.95-0.96.
In what follows we will only discuss
calculations using the SRC function $\hat{g}(r_r) = 1 - j_0 (q_cr_r)$ with
$q_c = 3.93 fm^{-1}$ \cite{to92}.

As discussed above and in Appendix B, the $\mu^-$ capture results depend on
the pseudoscalar coupling constant, $g_p$.  This dependence offers the
opportunity to extract a value for $g_p$ from a comparison of the $\beta^-$
and $\mu^-$ results. For the WBN interaction and WS
wave functions with the $1 - j_0(q_cr_r)$ SRC --- our preferred selection ---
the adopted value $\epsilon^\beta_{exp}$ = 1.61 $\pm$ 0.03 (which we assume to
yield $\epsilon^\mu_{exp}$ = 1.59 $\pm$ 0.04) and use of Eq. (22) leads to
$g_p$ = 7.4 $\pm$ 0.5 consistent with the PCAC value. Note that the
uncertainties assigned to $\epsilon_{exp}$ and $g_p$ reflect experiment only
and do not include any estimate of the uncertainty associated with the
theoretical analysis.

Finally we compare these results for the A = 16 mass region with our recent
results for the A = 132 \cite{wa92c} and A = 208 \cite{wa93c} regions.
As noted in Sec.
III.A the calculations in the heavier mass regions are further complicated by
the need to introduce core-polarization corrections.  These corrections
depend sensitively on the strength of the residual interaction used in their
evaluation and in particular on the strength of the tensor force which,
according to Brown and Rho \cite{br90}, is liable to medium modifications
similar to those proposed for the pion-decay constant \cite{br91,ku91}.  With a
weak tensor force, the enhancement factors deduced are
$\epsilon^\beta_{exp}$(A=132) = 1.82 $\pm$ 0.07 and
$\epsilon^\beta_{exp}$(A=208) = 1.79 $\pm$ 0.04. With strong tensor forces,
values of $\epsilon^\beta_{exp}$ some $10\%$ larger are
obtained.  Thus there is clearly more enhancement in heavy nuclei than in
light nuclei, which are characterized by our current value
$\epsilon_{exp}$(A=16) = 1.61 $\pm$ 0.03.  To quantify the mass dependence we
define a ratio,
\begin{equation}
r = {{\epsilon_{exp}(A=208)-1} \over {\epsilon_{exp}(A=16)-1}}, \label{rto}
\end{equation}
which for weak tensor forces has a value 1.30 $\pm$ 0.10 and for strong tensor
forces a value 1.44 $\pm$ 0.11.  The calculated value of this ratio from
meson-exchange models obtained by Towner \cite{to92} is 1.38, which is
approximately midway between
the experimental results obtained assuming weak and  strong tensor forces.
There is still some room for further sources of enhancement in the heavier
nuclei such as medium modification of the nucleon mass and of the pion-decay
constant used in the soft-pion approximation as proposed by Kubodera and Rho
\cite{ku91} on the basis of a scaling in the effective chiral Lagrangian
discussed by Brown and Rho \cite{br91}.  But the compelling need for such an
alternative explanation that was thought present a few years ago now appears
to be absent.

\acknowledgments

We thank S. Nozawa, K. Kubodera, and H. Ohtsubo for their generous and expert
help in understanding the details of their study of $^{16}$N $\leftrightarrow$
$^{16}$O, Ref. \cite{no86}. We especially thank H. Ohtsubo for providing us
with his computer program for solving the Dirac equation for the wave function
of a K-shell muon and for instructions in its use.
Research was supported in part by the U. S. Department of Energy under Contract
Nos. DE-AC02-76CH00016 with Brookhaven National Laboratory and in part by the
National Science Foundation under Grant No. PHY-90-17077 with Michigan State
University. BAB would also like to acknowledge support from the
Humboldt Foundation.


\appendix
\section{$\mu^-$ CAPTURE}

Nozawa, Kubodera, and Ohtsubo \cite{no86} --- referred to as NKO --- developed
an approach to $\mu^-$ capture on $^{16}O$, which can be used with numerical
solutions of the Dirac equation.  In this appendix we wish to make the
bridge between the $\mu^-$ capture  formalism developed by NKO and
numerical results found by combining their solution of the Dirac
equation with our shell-model wave functions.  In essence this means
we will give explicit expressions for the radial dependence of the
matrix elements, which will then be evaluated by numerical integration.

Our expression for the $\mu^-$ capture rate [Eq. (\ref{lam:2})] is
\begin{equation}
\Lambda_{\mu} = C_{R}{{G^2} \over {2\pi}}{{\omega^2} \over
{1 + \omega/M_f}}{{|\overline{\phi_{1s}(0)}|^2} \over {{\lambdabar_{Ce}}^2}}
\Biggl[ {{g_A(q^2)} \over {g_A(0)}} \Biggr]^2\!|M_0^{\mu}|^2.  \label{A1}
\end{equation}

\noindent In order to evaluate relativistic effects, NKO applied a
Foldy-Wouthuysen transformation first to order $1/M_N$ and then to
order $1/M^2_N$.  They showed that the terms of order $1/M^2_N$
contribute $\sim 2\%$ (constructively) to the $\mu^-$ capture  matrix
element $M^{\mu}_0$.  We will calculate $\Lambda_\mu$ to first order in
$1/M_N$ but include a multiplicative factor of $C_{R}$ (=1.04) in
Eq. (\ref{A1}) to compensate for the $1/M^2_N$ term in our evaluation of
$M^\mu_0$.

In Eq. (\ref{A1}) $|\overline{\phi_{1s}(0)}|^2$ is the
probability of finding a K-shell muon at the origin and is just $1/4\pi$
times the square of the large component of the muon wave function $G_{-
1}(r)$ evaluated at $r=0$.  For a point nucleus \cite{be82}
\begin{mathletters}
\label{A2}
\begin{equation}
G^{(p)}_{-1}(r) =
\Bigl[ {{(1+\gamma)\lambda} \over {\Gamma (1+2\gamma)}} \Bigr]^{1/2}
r^{-1} (2\lambda r)^{\gamma} e^{-\lambda r}, \qquad \label{A2a}
\end{equation}
\begin{equation}
F^{(p)}_{-1}(r) = - {{\alpha Z} \over {1+ \gamma}}G^{(p)}_{-1}(r),\label{A2b}
\end{equation}
\end{mathletters}
\noindent where $F_{-1}(r)$ is the smaller component of the muon wave
function and $\gamma = [1 - (\alpha Z)^2]^{1/2}$.  When appropriate, we use
the superscripts $(p)$ and $(g)$ on $G_{-1}$ and $F_{-1}$ to denote a point
nucleus and a nucleus with a Gaussian charge distribution, respectively.
In Eq. (A2)
\begin{equation}
\lambda = \Bigl[ (m^r_{\mu} + E_{\mu})(m^r_{\mu} - E_{\mu}) \Bigr]^{1/2},
\qquad E_{\mu} = m^r_{\mu} \gamma \label{A3}
\end{equation}
where $E_{\mu}$ is the energy eigenvalue for the K-shell for a point
nucleus.  The approximation often used for $G_{-1}^{(p)}(r)$ is
\begin{equation}
G_{-1}^{(p)}(r) \buildrel Z \rightarrow 0 \over \longrightarrow
2 (\alpha Zm^r_{\mu})^{3/2}.\label{A4}
\end{equation}
\noindent Thus we take the probability of finding the muon at the origin
(in the unit volume
$\mathchar'26\mkern-9mu\lambda_{Ce}
^3$) to be
\begin{equation}
{|\overline{\phi_{1s}(r)}|}^2_{r=0} =
{{1} \over {4\pi}} [G_{-1}(0)]^2 =
{{1} \over {\pi}} (\alpha Zm^r_{\mu})^3 R^2_Z,\label{A5}
\end{equation}
where $R_Z$ is a correction evaluated by solving the Dirac equation for a
realistic extended nuclear-charge distribution.

We use NKO's solution of the Dirac equation.  These authors used a Gaussian
charge distribution $\rho(r)$ given by
\begin{equation}
\rho(r) = \rho_0 \bigl[1 + a (r/r_0)^2 \bigr] e^{-(r/r_0)^2},\label{A6}
\end{equation}
\noindent with $r_0 = 1.83$ $fm$ and $a = 1.45$.  This charge distribution
gives $G_{-1}^{(g)}(0) = 0.91688$ and large and small components of the
muon wave function, which are well reproduced by the phenomenological power
expansion
\begin{equation}
G_{-1}^{(g)}(r) =
\sum^{10}_{n=0} a_nr^n, \qquad
F_{-1}^{(g)}(r) =
\sum^{10}_{n=0} b_nr^n,\label{A7}
\end{equation}
with the coefficients given in Table \ref{a1}.

The Foldy-Wouthuysen transformation to first order in $1/M_N$ leads to a
one-body operator for $\mu^-$ capture given by \cite{no86}
\begin{equation}
J_{\mu}^{1-body} = [i {\bf A} \cdot \hat{\bf r} L^+(r) +
A_0 {\cal {L}}^-(r)
+ i {\tilde {A}} {\cal {L}}^+(r)] \pmb{$\tau$} \label{A8}
\end{equation}
\noindent with radial functions
\begin{equation}
L^{\pm}(r) = {{1} \over {\sqrt 2}} [G_{-1}(r) j_1(\omega r)
\pm F_{-1}(r) j_0(\omega r)], \label{A9}
\end{equation}
\begin{equation}
{\cal L}^{\pm}(r) = {{1} \over {\sqrt 2}} [G_{-1}(r) j_0(\omega r)
\pm F_{-1}(r) j_1(\omega r)], \label{A10}
\end{equation}
\noindent and operators
\begin{equation}
i{\bf A} \cdot \hat{\bf r} = -g_A \pmb{$\sigma$} \cdot \hat{\bf r},
\label{A11}
\end{equation}
\begin{equation}
A_0 = ig_A \pmb{$\sigma$} \cdot {\bf P}/2M_N, \label{A12}
\end{equation}
\begin{equation}
\tilde{A} = m_{\mu} G_p \pmb{$\sigma$} \cdot {\bf k}/2M_N. \label{A13}
\end{equation}
\noindent Here
$j_0(\omega r)$ and
$j_1(\omega r)$ are spherical Bessel functions representing the neutrino wave
functions.  Note that {\bf P} and {\bf k} depend on the momenta of the
initial and final nucleons, {\bf P} = {\bf p}$_i + $ {\bf p}$_f$ and {\bf k}
= {\bf p}$_i$ $-$ {\bf p}$_f$, which occur in combination with radial
functions ${\cal L}(r)$.  In the Fourier transform to coordinate space
these momenta transform into derivative operators according to the
replacement rules:
\begin{equation}
{\bf P} {\cal L}(r) \rightarrow -i \hat{\bf r} {\partial \over \partial r}
{\cal L}(r) - 2i {\cal L}(r) \pmb{$\nabla$}, \label{A14}
\end{equation}
\begin{equation}
{\bf k} {\cal L}(r) \rightarrow i \hat{\bf r} {\partial \over \partial r}
{\cal L}(r). \label{A15}
\end{equation}
\noindent Then, upon evaluating the partial derivatives, we find
expressions for the full matrix element evaluated to order $1/M_N$.  In
terms of the operators defined in Eq. (1) and the normalization imposed by
the expression for the $\mu^-$ capture rate in Eq. (A1), the one-body
operator becomes
\begin{equation}
J_{\mu}^{1-body} = a_T^{\mu} M_0^T - a_S^{\mu} M_0^S\label{A16}
\end{equation}
with
\begin{equation}
a^\mu_T = {{1} \over {G_{-1}(0)}} \biggl[
G_{-1}(r)j_0(\omega r) -
F_{-1}(r)j_1(\omega r)\biggr], \label{A17}
\end{equation}
\begin{eqnarray}
a^\mu_S = {{1} \over {rG_{-1}(0)}} &&\biggl[
[g_{large} + \delta_{large}(r)] G_{-1}(r)j_1(\omega r) \label{A18} \\
&&\!\!\!+ ~[g_{small} + \delta_{small}(r)]F_{-1}(r)
j_0(\omega r)\biggl], \nonumber
\end{eqnarray}
\noindent where in our notation $M_0^T$ and $M_0^S$ are equivalent to NKO's
notation $\mbox{$-{1 \over M_N}$ \pmb{$\sigma$} $\cdot$ \pmb{$\nabla$}
\pmb{$\tau$}$_-$}$ and $\mbox{\pmb{$\sigma$} $\cdot$ {\bf r}
\pmb{$\tau$}$_-$}$.
In Eq. (A18)
\begin{displaymath}
g_{large} = {3 \over \omega} g_\mu= 1- {\omega \over 2M_N}
(g_p-1) = 0.6969, \nonumber 
\end{displaymath}
\begin{displaymath}
g_{small} = 1 + {1 \over 2M_N} \biggl[ 2m_\mu(g_p -1) +
\omega (g_p + 1)\biggr] = 2.0753, \nonumber 
\end{displaymath}
\begin{displaymath}
\delta_{large}(r) = {1 \over 2M_N} [V(r) + E_K](g_p + 1), \nonumber
\end{displaymath}
\begin{displaymath}
\delta_{small}(r) = {-1 \over 2M_N} \biggl[ [V(r) + E_K](g_p - 1) +
{4 \over r} (g_p + 1)\biggr], \nonumber 
\end{displaymath}
\noindent and $g_p = m_{\mu} G_p(q^2)/g_A(q^2)$ is the pseudoscalar
coupling constant.  In these expressions, $E_K$ is the $K$-shell $\mu^-$
binding energy [$\equiv (1-\gamma)m_{\mu}$ for a point nucleus], and $V(r)$
is a potential evaluated from the charge distribution used to calculate
$G_{-1}(r)$ and $F_{-1}(r)$. The potential is
\begin{equation}
V(r) = {1 \over r} \int^r_0 \rho(x)x^2dx + \int^\infty_r \rho(x)xdx
\label{A23}
\end{equation}
with
\begin{equation}
\int^\infty_0 \rho(r)dr = \alpha Z. \label{A24}
\end{equation}

The  approximate solution designated ``simple'' by NKO corresponds to the
limits
\begin{eqnarray}
&&G_{-1}(r) \rightarrow G_{-1}(0), \nonumber   \\
&&F_{-1}(r) \rightarrow 0,~~~~~~~~~~~~~~~~{\rm ``simple''~ solution}
\label{A25} \\  &&V(r) + E_K \rightarrow 0. \nonumber
\end{eqnarray}
We have arranged the terms in Eqs. (A17,A18) so that it is easily
apparent that in this limit Eq. (A16) reduces to
\begin{equation}
J^{1-body}_\mu (simple) = j_0(\omega r)M^T_0+g_{\mu}\frac{3}{\omega r}
j_1(\omega r) M^S_0. \label{1-body}
\end{equation}
%
\noindent The terms in $\delta_{small}(r)$ and $\delta_{large}(r)$ give
small contributions ($\le$ 2\%). As pointed out by NKO, the $g_{small}$
term gives a contribution of order 10\%.

\section{PSEUDOSCALAR COUPLING CONSTANT}

For semi-leptonic weak processes, the $V-A$ structure of the weak
interaction is modified by the induced weak currents that arise from
the presence of the strong interaction.  For the axial current one
common parameterization is
\begin{equation}
A_{\mu} = i \overline{u} (p_f)
[g_A(q^2) \gamma_{\mu} \gamma_5
- {{G_p(q^2)} \over {2M}}\ q_{\mu}\negthinspace\not\!{q}\gamma_5]
u(p_i) \case{\tau^a}{2} \label{B1}
\end{equation}
where the strong interactions have renormalized the axial-vector
coupling constant $g_A(q^2)$ from a bare value of unity and introduced
a pseudoscalar term with coupling constant $G_p(q^2)$.  Here
$\overline{u}(p_f)$ and $u(p_i)$ are spinors representing the final
and initial state nucleons, $q$ is the momentum transfer, and $\tau^a$ is the
Pauli isospin matrix with the superscript $a$ representing the Cartesian index
$x \pm iy$.  If,
additionally, a constraint is imposed on Eq. (B1), namely that the
axial current shall be partially conserved, the PCAC condition, then a
relation between $G_p(q^2)$ and $g_A(q^2)$ can be established.

One model for the weak axial current of a nucleon, that satisfies by
construction the PCAC condition, is one of meson dominance discussed
by Towner \cite{to92}.  In this model, the $g_A$ term in Eq. (B1)
originates in the axial current being mediated by the $A_1$ meson in
its interaction with a nucleon, while the $G_p$ term is mediated by
the $\pi$-meson.  The chiral Lagrangian that is used to describe the
meson-nucleon interaction preferentially chooses pseudovector coupling
for $\pi$NN vertices.  Hence there is a pseudovector form,
$q_{\mu}\negthinspace\not\!{q}\gamma_5$, to the $G_p$ term in Eq.
(B1).  However for on-mass-shell nucleons, the use of the Dirac
equation can transform this equation into one with pseudoscalar
coupling, $q_{\mu} \gamma_5$.  Using this meson dominance model,
Towner \cite{to92} obtains
\begin{equation}
G_p(q^2) = {{2M g_A(q^2)} \over {q^2 + m^2_{\pi}}} \left( 1 - {{m^2_{\pi}}
\over {m_A^2}}\right). \label{B2}
\end{equation}
\noindent The last factor, which tends to unity in the limit
$m_{\pi}/m_A \rightarrow 0$, is not usually present in the literature.
To get this result from the chiral Lagrangian the Weinberg relation is
used relating the $A_1$-meson mass to the $\rho$-meson mass:  $m_A^2 =
2m^2_{\rho}$.  With an experimental $\rho$-meson mass of $768.1$ MeV,
this leads to an $A_1$-meson mass of $m_A = 1086.3$ MeV.  On the other
hand, neutrino-nucleon scattering analysed using $g_A(q^2) = g_A(0)
m^2_A/(m_A^2 + q^2)$ deduce that $m_A \simeq 950$ MeV.  Thus there is
some small uncertainty on the precise value of $m_A$ to use in Eq.
(B2).  We will take $m_A = 950$ MeV and use for the nucleon and pion
masses their isospin averages to obtain for the dimensionless quantity
\begin{equation}
g_p \equiv {{m_{\mu} G_p(q^2)} \over {g_A(q^2)}} = {{2M m_{\mu}}
\over {q^2 + m^2_{\pi}}} \left(1 - {{m^2_{\pi}} \over {m_A^2}}\right)
 = 6.939\label{B3}
\end{equation}
\noindent for $q^2 = 0.8 m^2_{\mu}$.  This is the value recorded in
Table III.

The pseudoscalar coupling constant, $g_p$, is the least well measured
of all the nucleon weak-interaction coupling constants.  Its value in
Eq. (B3) typically represents a free-nucleon value and could be
modified when that nucleon is embedded in a nuclear medium.  The most
promising way to determine $g_p$ experimentally is in radiative $\mu^-$
capture \cite{Ar92}.  The branching ratio for radiative $\mu^-$ capture
relative to ordinary (non-radiative) $\mu^-$ capture is particularly
sensitive to $g_p$.  However the extraction of $g_p$ from nuclear
radiative $\mu^-$ capture measurements requires a model calculation of
the inclusive nuclear response function and this piece of the determination is
not yet under good control.  For example the most precise measurement
for the $^{16}O$ branching ratio \cite{Ar92} yields values of $g_p =
7.3 \pm 0.9$, $9.1 \pm 0.9$, and $13.6 \pm 1.9$ when compared with
three different recent calculations of the nuclear response.
Nevertheless there is a hint here that the value of $g_p$ in $^{16}O$
is larger than the PCAC value, Eq. (B3).  A similar statement can be
made for $^{12}C$.  For heavier nuclei, analysed using a Fermi-gas
model calculation of the nuclear response, the indication \cite{Ar92,Do88}
is that $g_p$ falls below the PCAC value and even quenches to
zero for nuclei as heavy as $Pb$.

For $^{16}O$, there is an alternative way to obtain $g_p$; namely, from the
first-forbidden $0^-$ to $0^+$ $\beta$ transition from $^{16}N$ and the
inverse $\mu^-$ capture on $^{16}O$, the reactions under study in this
paper.  Gagliardi {\it et al} \cite{ga83b} using the model
calculations of Towner and Khanna \cite{to81} deduce $g_p = 11 \pm 2$.
This result is model dependent so that our evaluation --- given in
Sec. \ref{sec:conclusions} results in the quite different
$g_p$ = 7.4 $\pm$ 0.5.
Because of the model dependence involved, we have decided in this paper to
quote results using the PCAC value of $g_p$, but give sufficient
details --- such as Eq. (\ref{gp}) --- that the results can easily be
modified for different values of $g_p$.


\begin{figure}
\caption[Figure 1]{The four $\Delta$J = 0 decays of interest to the present
study. Each level is labeled by $J^\pi$ and $E_x$ (in keV). Each $\beta^-$
branch is labeled by the percentage branching ratio (or decay rate) and
$\log$$f_0t$ value.}
\label{decay}
\end{figure}
\begin{figure}
\caption[Figure 2]{Comparison of partial WBN and experimental level schemes of
$^{16}$O. The experimental energies and spin-parity assignments are from Ref.
\protect\cite{aj82}. The theoretical levels were calculated with the
indicated downward shifts to the different n$\hbar\omega$ components relative
to the unshifted 0$\hbar\omega$ components. The centroid of the 1$\hbar\omega$
+ 3$\hbar\omega$ shifts was set so as to minimize the difference in excitation
energies for the yrast T = 1 odd-parity quartet. The other shifts were set as
discribed in the text.  The levels included in the figure are the lowest
five 0$^+$ T = 0 level, the odd-parity T = 1 quartet, and all T = 0 odd-parity
levels below 12-MeV excitation. There is no WBN counterpart for the known
predominantly 5$\hbar\omega$ experimental 1$^-$ and 3$^-$ states at 9585 and
11600 keV, respectively. The experimental 0$^+$ level at 11260 keV is
tentative.
\protect\cite{aj82}. We assume it does not exist.}
\label{16}
\end{figure}
\begin{figure}
\caption[Figure 3]{Experimental data for R0 and R2 $\beta^-$ decays of the
$0^-$ and $2^-$ states of $^{16}$N.  Each level is labeled by J$^\pi$ and
E$_x$ (in keV). The R0 decay is labeled by the decay rate and the R2 decays is
labeled by the total half-life and the individual branching ratios.}
\label{16N}
\end{figure}
\begin{figure}
\caption[Figure 4]{Schematic showing the contributions of n$\hbar\omega$
$\rightarrow$ (n$\pm$1)$\hbar\omega$ transitions to the $^{16}$N
$\leftrightarrow$ $^{16}$O \noindent R0 $M^T_0$ matrix element in the $0^-$
$\leftrightarrow$ $0^+_1$ transition (4a) and the R2 $M^{\textstyle z}_2$
matrix element in the $2^-$ $\leftrightarrow$ $0^+_1$ transition (4b). Both
are calculated with WS wave functions and the WBP
interaction. $\alpha_{in}$ and $\alpha_{fn}$ are the amplitudes of the various
$\hbar\omega$ components in the initial and final states, respectively.}
\label{schematic}
\end{figure}
\newpage
\onecolumn
\squeezetable
\mediumtext
\begin{table}
\caption[Table I]{Experimental data for $\Delta$J = 0 first-forbidden decays in
the A $\sim$ 16 region. $t_{1/2}$ is the total half-life and $b_r$ the
branching ratio for the indicated final state.  Q$_{\beta}$ is defined
in Eq. (\protect \ref{snp2}). ${\overline {C(W)}}$ is the average shape factor,
$\equiv$ $f/f_0$ [see Eq. (\protect \ref{f})], and is given by
${\overline {C(W)}}$ = $\sum_R$ B$^{(R)}_1$ where B$^{(R)}_1$ $\equiv$
$\vert M^\beta_R \vert^2$ is the rank R $\beta$ moment. For the $\case{1}{2}^+$
$\rightarrow$ $\case{1}{2}^-$ transitions R = 0,1 and the measured
$B^{(1)}_1/B^{(0)}_1$  listed in the sixth column is used to obtain $M^\beta_0$
from ${\overline {C(W)}}$. For the $0^+ \leftrightarrow$ $0^-$ transitions
this step is not necessary.}
\begin{tabular}{cccccccc}
Transition&$t_{1/2}$&$b_r$ & Q$_{\beta}$   & ${\overline {C(W)}}$$^{1/2}$ &
$B^{(1)}_1$/$B^{(0)}_1$& $M^\beta_0$& Ref. \\
           &  (s)   & (\%) &(keV)& ($fm$) &        & ($fm$)     &      \\
\tableline
&&&&&&& \\
$^{11}$Be($\case{1}/{2}^+$) $\rightarrow$ $^{11}$B($\case{1}/{2}^-$) &
\dec 13.81(8) &\dec 31.4(18) &\dec 9381.3(60) &\dec 14.4(4) &\dec $<$0.29 &
\dec 13.3(11) &\cite{mi82,wa82} \\
&&&&&&& \\
$^{15}$C($\case{1}/{2}^+$) $\rightarrow$ $^{15}$N($\case{1}/{2}^-$) &
\dec 2.449(5) &\dec 36.8(8) &\dec 9771.68(80) &\dec 32.8(4) &\dec 0.185(13) &
\dec 29.6(11) &\cite{wa85} \\
&&&&&&& \\
$^{16}$C($0^+$) $\rightarrow$ $^{16}$N($0^-$) &\dec 0.747(8) &\dec 0.68(10) &
\dec 7891.7(43) &\dec 13.4(9) &\dec 0.00 &\dec 13.4(9) &\cite{ga83a} \\
&&&&&&& \\
$^{16}$N($0^-$) $\rightarrow$ $^{16}$O($0^+$) &\dec 1.429(56) &\dec 100.0(0) &
\dec 10539.5(23) &\dec 58.4(11) &\dec 0.00 &\dec 58.4(11) &\cite{ga83b,ha85} \\
\end{tabular}
\label{data}
\end{table}
\squeezetable
\widetext
\begin{table}
\caption[Table II]{Summary of results for the neutron and proton separation
energies $S(n)$ and $S(p)$ for use with Woods-Saxon wave functions. All
energies in the last seven columns are in keV.}
\begin{tabular}{ccrrrrrrr}
Transition & j & $\Delta E_b(n)$ &$\Delta E_b(p)$ & $E_i$ & $E_f$ &
$\langle E_x \rangle$ & $S(n)$ & $S(p)$ \\
\tableline
&&&&&&&& \\
$^{16}$N$(0^-)$ $\rightarrow$ $^{16}$O$(0^+)$ &
$\case{1}/{2}$ & 2491 & 12128 &  120 &     0 &    0 & 2371 & 12128 \\
&&&&&&&& \\
                                                                    &
$\case{3}/{2}$ &      &       &      &       & 6324 & 8695 & 18452 \\
&&&&&&&& \\
$^{16}$C$(0^+)$ $\rightarrow$ $^{16}$N$(0^-)$ &
$\case{1}/{2}$ & 4251 & 11480 &    0 &   120 &  25 & 4276 & 11385 \\
&&&&&&&& \\
                                                                    &
$\case{3}/{2}$ &      &       &      &       & 4267 & 8518 & 15627 \\
&&&&&&&& \\
$^{15}$C$(\case{1}/{2}^+)$ $\rightarrow$ $^{15}$N$(\case{1}/{2}^-)$ &
$\case{1}/{2}$ & 1218 & 10208 &    0 &     0 &    0 & 1218 & 10208 \\
&&&&&&&& \\
                                                                    &
$\case{3}/{2}$ &      &       &      &       & 8746 & 9964 & 18954 \\
&&&&&&&& \\
$^{11}$Be$(\case{1}/{2}^+)$ $\rightarrow$ $^{11}$B$(\case{1}/{2}^-)$ &
$\case{1}/{2}$ &  504 & 11228 &    0 &  2125 &   70 &  574 &  9173 \\
&&&&&&&& \\
                                                                     &
$\case{3}/{2}$ &      &       &      &       & 7013 & 7517 & 16116 \\
&&&&&&&& \\
\end{tabular}
\label{spn}
\end{table}
\narrowtext
\begin{table}
\squeezetable
\caption[Table III]{Quantities needed in the evaluation of the $^{16}$N($0^-$)
$\leftrightarrow$  $^{16}$O($0^+$) $\beta^-$ and $\mu^-$ matrix elements and
decay rates.}
\begin{tabular}{@{\hspace{-0.2in}}c@{\hspace{-0.2in}}c@{\hspace{0.10in}}c
@{\hspace{0.02in}}c}
Quantity & Definition & Value & Ref. \\
\tableline
$\alpha$ & fine structure constant & 1/137.036 & \cite{pdg} \\
${\mathchar'26\mkern-9mu\lambda_{Ce}}$ & electron Compton wavelength &
386.159 $fm$ & \cite{pdg} \\
$r_u$ & uniform charge radius & 3.537 $fm$ & \cite{wi77} \\
$\xi$ & $\alpha Z/2r_u$ & 3.187 & \\
$\gamma$ & $[1 - (\alpha Z)^2]^{1/2}$ & 0.99829 &\cite{be82} \\
$M_N$   & $T_Z$-averaged nucleon mass & 1837.41 & \cite{pdg} \\
$m_\pi$ & $T_Z$-averaged pion    mass &  270.128 & \cite{pdg} \\
$m_\mu$ & muon mass  & 206.768 & \cite{pdg} \\
$m^r_\mu$ & reduced muon mass & 0.99296$m_\mu$ & \\
$G^2/2\pi^3$ & Weak-interaction factor ($sec^{-1}$)  &
 ~[8851(20)]$^{-1}$ & \cite{pdg} \\
$\omega$ & $m_\mu$ $-$ Q$_{\beta}$ & 186.144 & \cite{to81} \\
$q^2$ & $-m^2_\mu + 2m_\mu \omega$ & 0.80051$m^2_\mu$ & \cite{to81} \\
$q$   & four-momentum transfer & 184.998 & \cite{to81} \\
$g_A$        & $G_A/G$ [$\equiv$ $g_A(0)$]    & 1.261(4) & \cite{pdg} \\
$g_A(q^2)/g_A(0)$ & $\lbrack 1 + (q/M_N)^2 \rbrack^{-1}$ & 0.9900 & \\
Q$_{\beta}$ & $\beta$-decay Q value & 20.624(5) & \cite{wap88} \\
$f_0/
{\mathchar'26\mkern-9mu\lambda_{Ce}}^2
{}~$ & phase-space factor & 1.2574 & \cite{wi74} \\
$\frac{1}{\pi}(Z\alpha m^r_\mu)^3$ & See Eq. (\protect\ref{phi0})  &
548.10 & \\
${\cal R}_Z$ & See Eq. (\protect\ref{phi0})
 & 0.91688 & \cite{no86} \\
$g_p(q^2)/g_A(q^2)$ & PCAC value &6.939 & \\
$E_\mu$(point nucleus) & $\gamma m^r_\mu$ & 204.961 & \cite{be82} \\
$E_K$(point nucleus)   & $(1 - \gamma)m^r_\mu$ & 0.350 & \cite{be82} \\
$(1+\omega/M_f)^{-1}$ & recoil correction & 0.9937 & \\
$C_{R}$              & relativistic correction & 1.04 &\cite{no86} \\
\end{tabular}
\label{def}
\end{table}

\squeezetable
\mediumtext
\begin{table}
\caption[Table IV]{Soft-pion enhancement factors for simple transitions
encountered in $^{16}$N($0^-$) $\rightarrow$ $^{16}$O($0^+$)
calculated with HO wave functions ($\hbar\omega$ = 13.60 MeV) for two different
short-range correlation functions. WS results are also given for the
$\nu 1s_{1/2}$ $\rightarrow$ $\pi 0p_{1/2}$ and $\nu 0d_{3/2}$ $\rightarrow$
$\pi 0p_{3/2}$ transitions. Note that the last three transitions have
an $^{16}$O 0$\hbar\omega$ core.}
\begin{tabular}{llcccc}
Initial State & Final State &\multicolumn{2}{c}{$\epsilon^\beta_{mec}$}
&\multicolumn{2}{c}{$\epsilon^\mu_{mec}$} \\
              &             & $1 - j_0(3.93r_r)$ & $\theta(r_r - 0.71)$
& $1 - j_0(3.93r_r)$ & $\theta(r_r - 0.71)$ \\
\tableline
&&& \\
\multicolumn{6}{c}{1$\hbar\omega$ $\rightarrow$ 0$\hbar\omega$} \\
&&& \\
$\nu 1s_{1/2} \pi (0p_{1/2})^{-1}$ & $(0s)^4(0p)^{12}$
&\dec 1.549$^{\rm a}$ &\dec 1.488\tablenotemark[1]
&\dec 1.510$^{\rm a}$ &\dec 1.425\tablenotemark[1] \\
&&\dec 1.523$^{\rm b}$ &\dec 1.467\tablenotemark[2]
&\dec 1.509$^{\rm b}$ &\dec 1.427\tablenotemark[2] \\
$\nu 0d_{3/2} \pi (0p_{3/2})^{-1}$ & $(0s)^4(0p)^{12}$
&\dec 1.418$^{\rm a}$ &\dec 1.383\tablenotemark[1]
&\dec 1.449$^{\rm a}$ &\dec 1.387\tablenotemark[1] \\
&&\dec 1.428$^{\rm b}$ &\dec 1.379\tablenotemark[2]
&\dec 1.457$^{\rm b}$ &\dec 1.393\tablenotemark[2] \\
 \\
&&& \\
\multicolumn{6}{c}{1$\hbar\omega$ $\rightarrow$ 2$\hbar\omega$} \\
&&& \\
$\nu 1s_{1/2} \pi (0p_{1/2})^{-1}$ & $\nu (1s_{1/2}) \nu (0s_{1/2})^{-1}$
&\dec 1.405 &\dec 1.367 &\dec 1.417 &\dec 1.361 \\
$\nu 1s_{1/2} \pi (0p_{1/2})^{-1}$ & $\nu (1s_{1/2})^2 \pi (0p_{1/2})^{-2}$
&\dec 1.525 &\dec 1.458 &\dec 1.488 &\dec 1.406 \\
$\nu 0d_{3/2} \pi (0p_{3/2})^{-1}$ & $\nu (0d_{3/2})^2 \pi (0p_{3/2})^{-2}$
&\dec 1.331 &\dec 1.293 &\dec 1.356 &\dec 1.306 \\
&&& \\
\multicolumn{6}{c}{1s0d $\rightarrow$ 0f1p} \\
&&& \\
$0d_{5/2}$ & $0f_{5/2}$ &\dec 1.226 &\dec 1.201
&\dec 1.253 &\dec 1.212 \\
$0d_{3/2}$ & $1p_{3/2}$ &\dec 1.488 &\dec 1.430
&\dec 1.436 &\dec 1.360 \\
$1s_{1/2}$ & $1p_{1/2}$ &\dec 1.305 &\dec 1.264
&\dec 1.310 &\dec 1.248 \\
\end{tabular}
\tablenotetext[1]{HO results.}
\tablenotetext[2]{WS results.}
\label{ist}
\end{table}
\narrowtext
\begin{table}
\squeezetable
\caption[Table V]{Results obtained for restricted model spaces. All were
obtained with the Woods-Saxon potential.}
\begin{tabular}{llccc}
Transition & Model space & M$_0^T$ & M$_0^S$ \\
\tableline
$^{11}$Be($\case{1}/{2}^+$) $\rightarrow$ $^{11}$B($\case{1}/{2}^-$) &
 single-particle & \dec 34.3 & \dec $-$1.59 \\
                                                                     &
1$\hbar\omega$ $\rightarrow$ 0$\hbar\omega$
 & \dec 18.0 & \dec $-$0.97 \\
$^{15}$C($\case{1}/{2}^+$) $\rightarrow$ $^{15}$N($\case{1}/{2}^-$) &
 single-particle & \dec 47.3 & \dec $-$2.20 \\
                                                                     &
1$\hbar\omega$ $\rightarrow$ 0$\hbar\omega$
 & \dec 35.9 & \dec $-$1.76 \\
$^{16}$C($0^+$) $\rightarrow$ $^{16}$N($0^-$) &
 single-particle & \dec 81.1 & \dec $-$3.12 \\
                                                                     &
2$\hbar\omega$ $\rightarrow$ 1$\hbar\omega$
 & \dec 26.8 & \dec $-$1.22 \\

$^{16}$N($0^-$) $\rightarrow$ $^{16}$O($0^+$) &
 single-particle & \dec 81.1 & \dec $-$3.12 \\
                                                                     &
1$\hbar\omega$ $\rightarrow$ 0$\hbar\omega$
 & \dec 62.1 & \dec $-$2.70 \\
\end{tabular}
\label{simple}
\end{table}
\mediumtext
\begin{table}
\squeezetable
\caption[Table VI]{Results obtained from the (1+3)$\hbar\omega$ $\rightarrow$
(0+2+4)$\hbar\omega$ calculations for $^{16}$N($0^-$) $\rightarrow$
$^{16}$O($0^+$) R0 $\beta$ decay. The $\epsilon^\beta_{exp}$ were obtained
from Eq. (\protect\ref{allmth}) using the experimental value of M$_0^{\beta}$
= 58.4 $\pm$ 1.1. The $\epsilon^\beta_{mec}$ were obtained
using the calculated matrix element M$^{\beta}_{\pi}$ as discussed in
Sec. III.C with the [1$-$$j_0(3.93r_r)$] short-range correlation.}
\begin{tabular}{ccddddd}
Interaction & WS/HO & M$_0^T$ & a$^\beta_S$ & M$_0^S$
& $~~~~~\epsilon^{\beta}_{exp}$
& $~~\epsilon^{\beta}_{mec}$ \\
\tableline
  WBP & HO & 49.9 & 10.01 & $-$2.59 & 1.69(2) & 1.63 \\
  WBN & HO & 56.6 &  9.49 & $-$2.26 & 1.41(2) & 1.60 \\
  WBP & WS & 41.7 &  9.37 & $-$2.58 & 1.98(3) & 1.61 \\
  WBN & WS & 48.8 &  9.36 & $-$2.26 & 1.63(2) & 1.62 \\
\end{tabular}
\label{highbe}
\end{table}

\mediumtext
\begin{table}
\squeezetable
\caption[Table VII]{Results obtained from the (1+3)$\hbar\omega$ $\rightarrow$
(0+2+4)$\hbar\omega$ calculations for $^{16}$O($0^+$) $\rightarrow$
$^{16}$N($0^-$)  $\mu^-$ capture. The $\epsilon^\mu_{exp}$ were obtained from
Eq. (\protect\ref{allmth}) using the experimental value of M$_0^{\mu}$ =
113.5 $\pm$ 3.4. The $\epsilon^\mu_{mec}$ was obtained
using the calculated matrix element M$^{\mu}_{\pi}$ as discussed in
Sec. III.C with the [1$-$$j_0(3.93r_r)$] short-range correlation and the PCAC
value of g$_p$ of Table III.}
\begin{tabular}{ccdddd}
Interaction & WS/HO & a$_T^{\mu}$ M$_0^T$
& a$_S^{\mu}$ M$_0^S$
& $~~~~~\epsilon^{\mu}_{exp}$
& $~~\epsilon^{\mu}_{mec}$ \\
\tableline
 WBP & HO & 46.5 & 65.9 & 1.02(7) & 1.60 \\
 WBN & HO & 51.5 & 57.2 & 1.09(7) & 1.59 \\
 WBP & WS & 36.8 & 56.9 & 1.54(9) & 1.61 \\
 WBN & WS & 41.8 & 48.8 & 1.55(8) & 1.59 \\
\end{tabular}
\label{highmu}
\end{table}
\narrowtext
\begin{table}
\squeezetable
\caption[Table VIII]{Results obtained from the (1+3)$\hbar\omega$ $\rightarrow$
(0+2+4)$\hbar\omega$ calculations for $^{16}$N($2^-$) $\rightarrow$
$^{16}$O($0^+_n$) R2 $\beta$ decay.}
\begin{tabular}{cdccd}
n & ~~Exp & Interaction & WS/HO & M$^{\textstyle z}_2(0^+_n)$  \\
\tableline
1 & 3.04(2)  &  WBP & HO & 2.91 \\
  &          &  WBN & HO & 2.61 \\
  &          &  WBP & WS & 2.93 \\
  &          &  WBN & WS & 2.63 \\
\tableline
2 & 1.09(18) &  WBP & HO & 0.96 \\
  &          &  WBN & HO & 1.03 \\
  &          &  WBP & WS & 0.96 \\
  &          &  WBN & WS & 1.03 \\
\end{tabular}
\label{highR2}
\end{table}
\squeezetable
\mediumtext
\begin{table}
\caption[Table IX]{WBN $^{16}$N($0^-$) $\rightarrow$ $^{16}$O($0^+$) results
for the $D^{(1)}_0(j_ij_f)$ and matrix elements of
Eq. (\protect\ref{mar}) calculated with WS wave functions for the dominant line
five and HO wave functions for the other entries as discussed in the text.}
\begin{tabular}{llrrrrr}
{}~~$\nu j_i$ &~~$\pi j_f$ &$D^{(1)}_0(j_ij_f)$ & $M^\beta_S(j_ij_f)$ &
${\cal M }^\beta_S(j_ij_f)$ &$M^\beta_T(j_ij_f)$ &${\cal M}^\beta_T(j_ij_f)$ \\
\tableline
  0p 1/2 &0s 1/2 &    0.0019 &    3.6473 &    0.0069 &      $-$105.7247 &
    $-$0.1988 \\
  1p 1/2 &0s 1/2 &   $-$0.0016 &    0.0000 &    0.0000 &      $-$0.0187 &
    $-$0.0000 \\
  0d 3/2 &0p 3/2 &    0.0504 &    6.9739 &    0.3515 &      $-$184.9045 &
    $-$9.3192 \\
  0s 1/2 &0p 1/2 &    0.0211 &    3.6139 &    0.0762 &         105.8366 &
      2.2310 \\
  1s 1/2 &0p 1/2 &    0.7857 &   $-$3.1164 &   $-$2.4486 &      73.0472 &
     57.3932 \\
  0f 5/2 &0d 5/2 &   $-$0.0062 &   10.1179 &   $-$0.0627 &  $-$269.3061 &
      1.6697 \\
  0p 3/2 &0d 3/2 &   $-$0.0424 &    6.9820 &   $-$0.2963 &     185.8391 &
    $-$7.8870 \\
  1p 3/2 &0d 3/2 &    0.0037 &   $-$4.4158 &   $-$0.0162 &     117.5352 &
      0.4302 \\
  0p 1/2 &1s 1/2 &   $-$0.0243 &   $-$3.1224 &    0.0759 &   $-$83.1269 &
      2.0216 \\
  1p 1/2 &1s 1/2 &   $-$0.0036 &    4.9370 &   $-$0.0178 &  $-$131.4311 &
      0.4732 \\
  0d 5/2 &0f 5/2 &   $-$0.0003 &   10.1179 &   $-$0.0030 &     269.3061 &
    $-$0.0808 \\
  0d 3/2 &1p 3/2 &    0.0013 &   $-$4.4158 &   $-$0.0057 &  $-$117.5352 &
    $-$0.1516 \\
  0s 1/2 &1p 1/2 &    0.0000 &    0.0000 &    0.0000 &           0.0187 &
    $-$0.0000 \\
  1s 1/2 &1p 1/2 &    0.0169 &    4.9370 &    0.0834 &         131.4311 &
      2.2199 \\
         &       &   Total   &           &   $-$2.2564 &                &
   48.8014 \\
\end{tabular}
\label{calm}
\end{table}
\narrowtext
\begin{table}
\squeezetable
\caption[Table X]{The power-series coefficients of Eq. (\protect\ref{A7})
that reproduce the Dirac wave function for the charge distribution of
Eq. (\protect\ref{A6}) with an absolute error $< 5\times10^{-6}$ for r
$< 10~fm$. The numbers in square brackets are powers of 10.}
\begin{tabular}{crccc}
&n & ~~~~~~~$a_n$ & ~~~~~~~$b_n$ &  \\
\tableline
&0 &\dec $+$ 9.16882[$-$01] &\dec $+$ 0.00000[$-$00] &\\
&1 &\dec $+$ 3.56365[$-$06] &\dec $-$ 8.29499[$-$03] &\\
&2 &\dec $-$ 4.51773[$-$03] &\dec $+$ 3.49933[$-$04] &\\
&3 &\dec $+$ 1.06662[$-$04] &\dec $-$ 3.02963[$-$04] &\\
&4 &\dec $-$ 5.80056[$-$05] &\dec $+$ 3.88561[$-$04] &\\
&5 &\dec $+$ 7.30971[$-$05] &\dec $-$ 1.40780[$-$04] &\\
&6 &\dec $-$ 2.20053[$-$05] &\dec $+$ 2.59785[$-$05] &\\
&7 &\dec $+$ 3.29970[$-$06] &\dec $-$ 2.77866[$-$06] &\\
&8 &\dec $-$ 2.76786[$-$07] &\dec $+$ 1.73880[$-$07] &\\
&9 &\dec $+$ 1.24699[$-$08] &\dec $-$ 5.87474[$-$09] &\\
\end{tabular}
\label{a1}
\end{table}

\begin{references}
\bibitem {wa91a} E. K. Warburton, Phys. Rev. C {\bf 44}, 1024 (1991).
\bibitem {wa92c} E. K. Warburton and I. S. Towner, Phys. Lett. B {\bf 294},
         1 (1992).
\bibitem {wa93c} E. K. Warburton and I. S. Towner, Phys. Rpts.,
         {\em to be published}.
\bibitem {mi82} D. J. Millener, D. E. Alburger, E. K. Warburton, and
         D. H. Wilkinson, Phys. Rev. C {\bf 26}, 1167 (1982).
\bibitem {wa82} E. K. Warburton, D. E. Alburger, and D. H. Wilkinson,
         Phys. Rev. C {\bf 26}, 1186 (1982).
\bibitem {wa85} E. K. Warburton, D. E. Alburger, and D. J. Millener, Phys. Rev.
         C {\bf 29}, 2281 (1984).
\bibitem {ga83a} C. A. Gagliardi, G. T. Garvey, N. Jarmie, and R. G. H.
         Robertson, Phys. Rev. C {\bf 27}, 1353 (1983).
\bibitem {ga83b} C. A. Gagliardi, G. T. Garvey, J. R. Wrobel, and
         S. J. Freedman, Phys. Rev. C {\bf 28}, 2423 (1983).
\bibitem {ha85} L. A. Hamil, L. Lessard, H. Jeremie, and J. Chauvin, Z. Phys.
         A {\bf 321}, 439 (1985);  H. Heath and G. T. Garvey, Phys. Rev. C
         {\bf 31}, 2190 (1985);  T. Minamisono, K. Takeyama, T. Ishigai,
         H. Takeshima, Y. Nojiri, and K. Asahi, Phys. Lett. {\bf 130B}, 1
         (1983);  L. Palffy {\it et al.}, Phys. Rev. Lett. {\bf 34}, 212
         (1975).
\bibitem {wa91} E. K. Warburton, Phys. Rev. Lett. {\bf 66}, 1823 (1991);
                Phys. Rev. C {\bf 44}, 233 (1991).
\bibitem {be82} H. Behrens and W. B\"uhring, {\it Electron Radial Wave
         Functions and Nuclear Beta-Decay}, Clarendon, Oxford (1982).
\bibitem {br66} G. E. Brown and A. M. Green, Nucl. Phys. {\bf 75}, 401 (1966).
\bibitem {wa92b} E. K. Warburton, B. A. Brown and D. J.
         Millener, Phys. Letts. {\bf B293}, 7 (1992).
\bibitem {br84}  B. A. Brown, A. Etchegoyen, W.D.M. Rae, and N. S.
         Godwin, OXBASH, 1984 ({\em unpublished}).
\bibitem {gl74} D. H. Glockner and R. D. Lawson, Phys. Lett. {\bf 53B},313
         (1974).
\bibitem {wa92a} E. K. Warburton and B. A. Brown, Phys. Rev. C {\bf 46}, 923
         (1992).
\bibitem {wa90a} E. K. Warburton, J. A. Becker, and B. A. Brown, Phys. Rev. C
         {\bf 41}, 1147 (1990).
\bibitem {ho85} A. Hosaka, K.-I. Kubo, and H. Toki,  Nucl. \ Phys.
         {\bf A244}, 76 (1985).
\bibitem {wa89} E. K. Warburton and D. J. Millener, Phys. Rev. C {\bf 39},
         1120 (1989).
\bibitem {ha90} W. C. Haxton and C. Johnson, Phys. Rev. Lett. {\bf 65},
                1325 (1990).
\bibitem {br93} B. A. Brown and E. K. Warburton, {\em unpublished}.
\bibitem{bo87} R. Machleidt, K. Holinde, and Ch. Elster, Phys. Rep. {\bf 149},
         1 (1987); R. Machleidt, Adv. Nucl. Phys. {\bf 19}, 189 (1989).
\bibitem {kuo93} T. T. S. Kuo, {\em private communication}.
\bibitem {wa88a} E. K. Warburton, J. A. Becker, B. A. Brown, and D. J.
         Millener, Anns. Phys. (N.Y.) {\bf 187}, 471 (1988).
\bibitem {footnote2} We use Edmonds' convention \cite{ed57} for reduced
         matrix elements, triple bars denote matrix elements reduced in both J
         and T, radial wave functions are positive at the origin, there is no
         $i^l$ with the spherical harmonic, and the order of coupling is
         {\boldmath ${l\times s}$}.
\bibitem {ed57} A. R. Edmonds, {\em Angular Momentum in Quantum Mechanics,}
         (Princeton University Press, 1957).
\bibitem {footnote4} In ``normal'' units $M^T_0$ is in (MeV-fm)$^{-1}$. To
         convert to natural units multiply by $m_e
\mathchar'26\mkern-9mu\lambda_{Ce}~$.
         Our convention of scaling matrix elements by
$\mathchar'26\mkern-9mu\lambda_{Ce}~$
         results in multiplication by $m_e
{\mathchar'26\mkern-9mu\lambda_{Ce}}^2
$.
\bibitem {de87} H. de Vries, C. W. de Jager, and C. de Vries, At. Data Nucl.
         Data Tables {\bf 36}, 495 (1987).
\bibitem {ba64} R. K. Bansal and J. B. French, Phys. Lett. {\bf 11}, 145
(1964).
\bibitem {mi85} D. J. Millener and E. K. Warburton, in {\em Nuclear Shell
   Models} (M. Vallieres and B. H. Wildenthal, Eds.), p. 365, World-Scientific,
    Singapore (1985).
\bibitem {no86} S. Nozawa, K. Kubodera, and H. Ohtsubo, Nucl. Phys. {\bf A453},
         645 (1986).
\bibitem {pdg}  Particle Data Group, Phys. Rev. D {\bf 45}, S1 (1992).
\bibitem {wi77} D. H. Wilkinson in {\em Ecole d'Et\'e de Physique
         Th\'eorique, Session XXX}, eds. R. Balian, M. Rho, and G. Ripka
         (Amsterdam, North-Holland) pp 879-1017 (1977).
\bibitem {to81} I. S. Towner and F. C. Khanna, Nucl. Phys. {\bf A372}, 331
         (1981).
\bibitem {wap88} A. H. Wapstra, G. Audi, and R. Hoekstra, At. Data Nucl. Data
         Tables {\bf 39}, 281 (1988); Midstream update of the mass table,
         private communication from G. Audi (1990).
\bibitem {wi74} D. H. Wilkinson and B. E. F. Macefield, Nucl. Phys.
         {\bf A232}, 1 (1974).
\bibitem {to86} I. S. Towner, Annu. Rev. Nucl. Part. Sci. {\bf 36}, 115 (1986).
\bibitem {footnote5} In the previous treatments of Towner and Khanna
\cite{to81}
         and Haxton and Johnson \cite{ha90}, an approximate expression,
         $a^\beta_S = 0.932[\case{1}/{3}(Q_\beta + 1) + \xi]$, was used rather
         than the accurate formulation of Ref. \cite{be82} given in
         Eq. (\ref{as}). Note that this approximation does not allow
         for the quite different contributions of the two terms in different
         transitions and thus is specific to this $^{16}$N decay. Even so it
         does not allow for the state-dependence of $r^\beta_S$. It should be
         noted that in previous treatments \cite{wa91,wa88a}
         $r^\beta_S$ was labeled $r^\prime_w$.
\bibitem {to92} I. S. Towner, Nucl. Phys. {\bf A542}, 631 (1992).
\bibitem {ku78} K. Kubodera, J. Delorme, and M. Rho, Phys. Rev. Lett.
      {\bf 40}, 755 (1978).
\bibitem {gu78} P. Guichon, M. Giffon, J. Joesph, R. Laverri\`ere, and C.
      Samour, Z. Phys. A {\bf 285}, 183 (1978); P. Guichon, M. Giffon, and
      C. Samour, Phys. Letts. {\bf 74B}, 15 (1978).
\bibitem {br80} B. A. Brown, W. A. Richter, and N. S. Godwin, Phys. Rev. Lett.
         {\bf 45}, 1681 (1980).
\bibitem {footnote6} This result was given on p340 of Ref. \cite{to81}, however
         in that equation the sign of the last term was incorrect.
\bibitem {wa68} E. K. Warburton, W. R. Harris, and D. E. Alburger, Phys. Rev.
         {\bf 175}, 1275 (1968).
\bibitem {wa88b}E. K. Warburton, in {\em Interactions and Structures in
Nuclei},
         Edited by R. Blin-Stoyle and W. Hamilton, (Adam Hilger, Bristol, GB,
         1988), pp. 81-88.
\bibitem {footnote7} It is difficult to make a detailed comparison between the
         results of Ref. \cite{ha90} and the present ones because various
         approximations were made in Ref. \cite{ha90} in the expressions
         for $\Lambda_\mu$ and $\Lambda_\beta$ and not enough detail is given
         to separate the one-body and two-body contibutions.
\bibitem {ah85} E. G. Adelberger and W. C. Haxton,
         Annu. Rev. Nucl. Part. Sci. {\bf 35}, 501 (1985).
\bibitem {br90}  G.E. Brown and M. Rho,  Phys. Lett. {\bf B237}, 3 (1990).
\bibitem {br91}  G.E. Brown and M. Rho,  Phys. Rev. Lett. {\bf 66}, 2720
         (1991).
\bibitem {ku91}  K. Kubodera and M. Rho,  Phys. Rev. Lett. {\bf 67}, 3479
         (1991).
\bibitem {Ar92}  D. S. Armstrong {\it et al.}, Phys. Rev. C {\bf 46}, 1094
         (1992)
\bibitem {Do88}  M. D$\ddot{o}$beli {\it et al.}, Phys. Rev. C {\bf 37},
  1633 (1988)
\bibitem {aj82} F. Ajzenberg-Selove, Nucl. Phys. {\bf A375}, 1 (1982).
\end{references}
\end{document}